\documentclass{article}

\usepackage[english]{babel}

\usepackage[letterpaper,top=2cm,bottom=2cm,marginparwidth=1.75cm]{geometry}

\usepackage[utf8]{inputenc}
\usepackage{amsmath}

\usepackage{graphicx}

\usepackage{booktabs}
\usepackage{longtable}
\usepackage{float}
\usepackage{multirow}

\usepackage{authblk}
\usepackage{titlesec}

\usepackage{csquotes}

\usepackage{subcaption}

\usepackage{soul}

\usepackage[colorlinks=true, allcolors=blue]{hyperref}

\usepackage[dvipsnames]{xcolor} 
\usepackage[normalem]{ulem}
\titlelabel{\thetitle.\quad}

\usepackage{siunitx}      
\providecommand{\keywords}[1]
{
  \small	
  \textbf{\textit{Keywords---}} #1
}

\title{Bridging classical and quantum interpretation of chemical state analysis by XPS/HAXPES to resolve short-range order in amorphous alumina films}

\author[1,2,3,4]{Simon Gramatte}
\author[5,6]{Xing Wang}
\author[7]{Michael Alejandro Hernández Bertrán}
\author[3]{Claudia Cancellieri}
\author[5,6]{Giovanni Pizzi}
\author[7]{Deborah Prezzi}
\author[6]{Iurii Timrov}
\author[2]{Olivier Politano}
\author[8]{Ivo Utke}
\author[3]{Lars P.H. Jeurgens}
\author[1,4,*]{Vladyslav Turlo}

\affil[1]{Laboratory for Advanced Materials Processing, Empa - Swiss Federal Laboratories for Materials Science and Technology, Feuerwerkerstrasse 39, 3602 Thun, Switzerland}
\affil[2]{Laboratoire Interdisciplinaire Carnot de Bourgogne, UMR 6303 CNRS-Université de Bourgogne, 9 Avenue A. Savary, 21078 Dijon Cedex, France}
\affil[3]{Laboratory for Joining Technologies and Corrosion, Empa - Swiss Federal Laboratories for Materials Science and Technology, Ueberlandstrasse 129, 8600 Duebendorf, Switzerland}
\affil[4]{National Centre for Computational Design and Discovery of Novel Materials (MARVEL), Empa, Thun, Switzerland}
\affil[5]{National Centre for Computational Design and Discovery of Novel Materials (MARVEL), 5232 Villigen PSI, Switzerland}
\affil[6]{PSI Center for Scientific Computing, Theory, and Data, 5232 Villigen PSI, Switzerland}
\affil[7]{Nanoscience Institute – National Research Council (CNR-NANO), I-41125 Modena, Italy}

\affil[8]{Laboratory for Mechanics of Materials and Nanostructures, Empa - Swiss Federal Laboratories for Materials Science and Technology, Feuerwerkerstrasse 39, 3602 Thun, Switzerland}
\affil[*]{Corresponding author: Vladyslav Turlo, vladyslav.turlo@empa.ch}

\begin{document}
\maketitle

\begin{abstract}
Probing the local structure and chemistry of wide-bandgap amorphous oxide thin films remains challenging due to the limitations of lab‑based spectroscopy. This work integrates X-ray photoelectron spectroscopy (XPS), hard X-ray photoemission spectroscopy (HAXPES), molecular dynamics simulations using machine-learning interatomic potentials, density-functional theory (DFT) calculations, and classical electrostatic modeling of final-state core-ionization effects in Al atoms to uncover the structure and chemistry of amorphous alumina polymorphs made with atomic layer deposition (ALD). DFT calculations using the $\Delta$Kohn-Sham method supported the interpretation of final-state effects and validated electrostatic model assumptions. Shifts in the measured Auger parameter were interpreted as extra-atomic relaxation energies, revealing sensitivity to the local coordination environment. Structural disorder and thermal fluctuations were found to govern the distribution of extra-atomic relaxation energies, suggesting that cryo-XPS can isolate and reveal intrinsic structural building blocks of amorphous oxides. Simulated heating and annealing demonstrated that Auger parameter shifts can serve as indicators of phase decomposition in H-supersaturated ALD amorphous alumina. These findings provide a pathway for comprehensive interpretation and predictive modeling of XPS spectra in amorphous wide-bandgap oxides.  
\end{abstract}


\textbf{\textit{Highlights}}

\begin{itemize}
\item Combined XPS/HAXPES and electrostatic modeling reveal coordination- and ligand-specific Auger parameter shifts in hydrogenated amorphous alumina.
\item Ligand polarizabilities of O$^{2-}$ and OH$^-$ are quantified by fitting experimental Auger shifts using Bayesian optimization.
\item $\Delta$Kohn-Sham DFT calculations capture metal-insulator core-level shifts and support interpretation of final-state effects.
\item Structural disorder governs the distribution of extra-atomic relaxation energies, suggesting cryo-XPS as a tool for probing local bonding.
\item Simulated annealing demonstrates the potential of Auger parameter shifts as sensitive indicators of thermal phase decomposition.
\end{itemize}

\begin{figure}[H]
    \centering
    \includegraphics[width=0.7\linewidth]{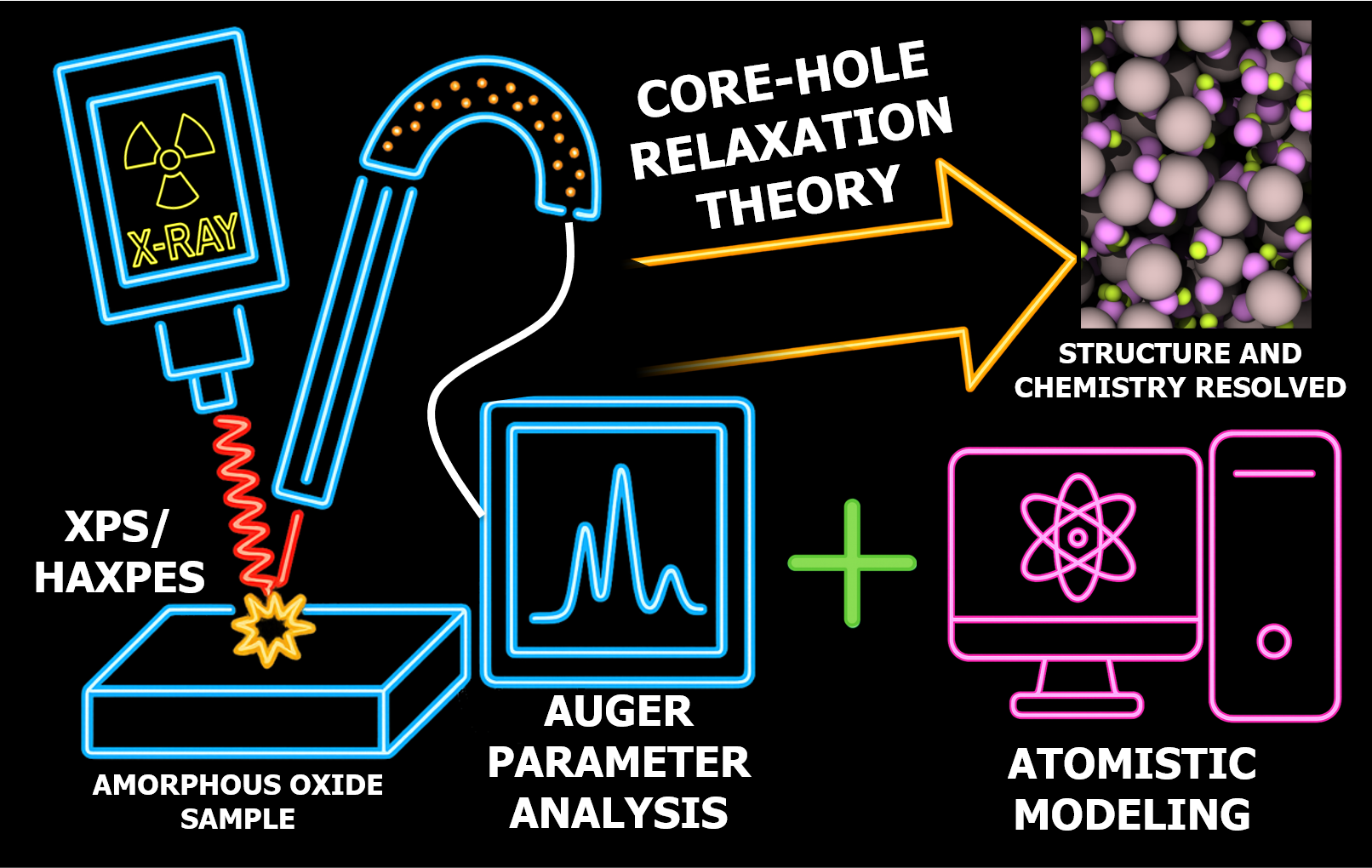}
    \label{fig:graphical}
\end{figure}

\pagebreak

\section{Introduction}
\label{Sec:introduction}

Subtle changes in local chemical states and short-range order in \textit{amorphous} oxides, such as those induced by light element impurities like hydrogen and/or by crystallization, are difficult to resolve with most analytical techniques. Among the few suitable methods are solid-state Nuclear Magnetic Resonance (NMR) \cite{Lee2009}, X-ray Adsorption Near-Edge and Extended Fine Structure Spectroscopy (XANES/EXAFS) \cite{Holgado2000}, Radial Distribution Function (RDF) analysis (e.g., by XRD or TEM) \cite{Lamparter1997,Terban2022}, and X-ray Photoelectron Spectroscopy (XPS) \cite{Jeurgens2000,Snijders2005,Cancellieri2024}. NMR and RDF are bulk-sensitive techniques that are not suitable for investigating films thinner than a few tenths of a nanometer. Quantitative XANES/EXAFS analyses typically require carefully selected crystalline reference materials for spectral deconvolution, yet may still fail to identify specific nearest-neighbour coordination spheres (NNCS) if their signatures are not present in the reference phases, such as for 5-fold or distorted 6-fold coordination spheres in amorphous alumina \cite{Gramatte2023}. High-resolution (HR) cross-sectional TEM, although applicable, is a destructive technique prone to electron-beam induced crystallization and/or radiation damage, especially in defective (hydr)oxide thin films in the absence of efficient sample cooling \cite{Reichel2007}. In contrast, soft and hard X-ray Photoelectron Spectroscopy (XPS and HAXPES, respectively) stand out as unique, powerful tools for probing local chemical states of oxide thin films, with probing depths that can reach about 6 nm for soft Al-K$\alpha$ (or Mg-K$\alpha$) and up to about 20-30 nm for hard Cr-K$\alpha$  radiation \cite{Jeurgens2024, Watts2024, Genz2024}. 

In our previous work~\cite{Gramatte2025}, we successfully reproduced the experimental structures and densities of atomic-layer-deposited (ALD) amorphous alumina polymorphs with varying hydrogen content by using atomistic simulations based on foundational neural network interatomic potentials. Bader charge analysis of the constituents (i.e., Al, O, and H atoms) in both amorphous and crystalline structures revealed distinct local chemical states that are largely independent of the oxide density and H content. These theoretical findings align with the established picture of the amorphous alumina structure as a randomly interconnected network of corner-sharing [AlO$_{n}$]-polyhedra~\cite{Snijders2005, Gramatte2022} ($n$ is the coordination of Al atom in the first neighbor shell), which defines its short-range order and gives it the characteristic bond flexibility (i.e."ductility")~\cite{Revesz1981, Zachariasen1932, Jeurgens2000,Frankberg2019}. 
During the ALD process, hydrogen atoms are incorporated in the amorphous alumina films and tend to form covalent bonds with the O ligands in the interconnected [AlO$_{n}$] polyhedra, with the effective replacement of some O atoms with hydroxyl groups in the NNCS of the Al cations. Based on these fundamental findings, the shifts in the Al Auger parameter corresponding to H incorporation measured by XPS/HAXPES \cite{Cancellieri2024} could be accurately predicted by assigning distinct polarizabilities to the O and OH ligands surrounding the core-ionized Al cations ~\cite{Gramatte2025}. However, in our previous study ~\cite{Gramatte2025}, only time- and structure-averaged properties, such as coordination numbers, bond lengths, geometric factors, and ligand fractions, were computed to comply with the simplified electrostatic model originally proposed for crystalline compounds \cite{Moretti1998}. The present study extends that work by adopting a refined electrostatic model that allows for the explicit resolution of the contribution of each [Al(O)$_{n-m}$(OH)$_{m}$] building block to the variance of the extra-atomic relaxation energy as a function of hydrogen content and temperature ($m$ represents the number of OH ligands in the first neighbor shell of Al atom).

To enable predictive modeling, the final-state effects (i.e., those induced by core holes) in amorphous oxides must be understood from first principles.  The most common approaches for the simulation of core-level spectra in the literature, i.e., $\Delta$Self-Consistent Field ($\Delta$SCF) and $\Delta$Kohm-Sham ($\Delta$KS) \cite{Kahk2019,Walter2016}, approximate the effect of an excited core-hole within density-functional theory (DFT), enabling the prediction of photoelectron binding energies. More advanced first principles methods, such as time-dependent DFT \cite{Besley2020}, multiconfigurational wave function and coupled cluster \cite{Liu2019b,Vidal2020}, and many-body perturbation theory~\cite{Golze2018,Golze2019a}, indeed offer higher accuracy but require orders of magnitude more computational resources than $\Delta$SCF/$\Delta$KS, particularly for amorphous systems requiring relatively large simulation cells to be properly described. While $\Delta$KS inherits some limitations from the underlying DFT functional, such as band-gap underestimation and ambiguities in the energy reference for insulators, it remains computationally efficient and sufficiently accurate to capture trends in electron density rearrangements and final-state screening following core-hole creation, which are the quantities needed here to investigate core-hole relaxation mechanisms in alumina. To our knowledge, this work is the first attempt to directly link the classical electrostatic theory of extra-atomic relaxation with the quantum mechanical calculation of excited, charged states, providing a valuable framework for interpreting experimentally observed core-level shifts.

\section{Core-hole relaxation theory}
\subsection{General formulation}
\begin{figure}[H]
     \centering
     \includegraphics[width=0.8\textwidth]{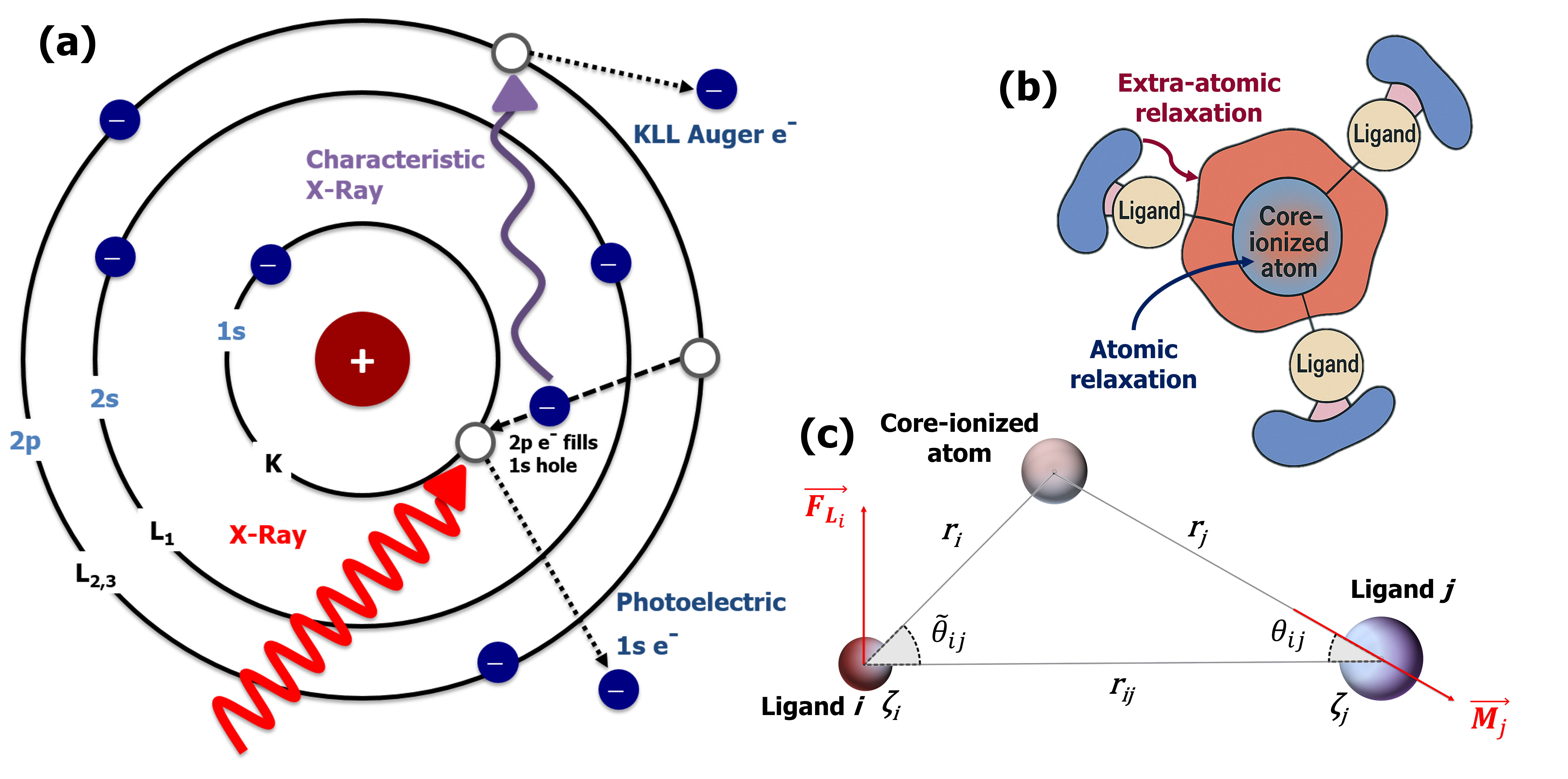}
     \caption{(a) Visualization of the KL$_{23}$L$_{23}$ Auger process (where K denotes the 1s shell, L$_{2}$ denotes the $2p\tfrac{1}{2}$ orbital and L$_{3}$ denotes the $2p\tfrac{3}{2}$ orbital): In a first step, an incoming photon excites a 1s core-level photoelectron. The binding energy of this photoelectron is determined by the difference between the incident X-ray energy and the detected kinetic energy of the emitted photoelectron ($E_{\mathrm{b}}^\mathrm{PE} = h\nu - E_{\mathrm{k}}^\mathrm{PE}$). The created core hole is then filled by an electron from a higher orbital; this transition releases energy, which ejects a second electron from another orbital, known as the Auger electron. The Auger electron transition involves three electronic orbitals: the core orbital from which the photoelectron is ejected, the higher-level orbital that fills the core hole, and another higher-level orbital from which the Auger electron is ejected.  (b) Schematic representation of extra-atomic screening of a core-ionized atom, as governed by the electronic polarizability of its neighboring ligands, where charge depletion (accumulation) is depicted in blue (red).
     (c) A diagram illustrating the electrostatic approach for the calculation of polarization energy. The model envisages a plane encapsulating the atom with a core hole alongside two adjacent ligands, labeled $i$ and $j$. According to electrostatic theory, the electric field at ligand $i$ is attributed to the induced dipoles on ligand $j$ ($M_j$) and the local field at ligand $i$. }
     \label{fig:Polarizability}
 \end{figure}

Recent advances in instrumentation, microelectronics, and detector technologies have enabled routine hard X-ray photoemission spectroscopy (HAXPES) measurements, both at synchrotrons and in the laboratory, using a combination of Al-K$\alpha$, Ag-K$\alpha$, Cr-K$\alpha$, and/or Ga-K$\alpha$ X-ray sources \cite{Watts2024}. The higher photon energies used in HAXPES produce photoelectrons with greater kinetic energies, thus enabling larger probing depths, up to 20–30 nm, for chemical state analysis \cite{Jeurgens2024}. Moreover, photoemission of deep core-level electrons (e.g., from the 1s shell) by hard X-rays can trigger sharp and intense core–core–core Auger transitions, particularly KL$_{23}$L$_{23}$ (see Fig. \ref{fig:Polarizability}a), which can be exploited for local chemical state analysis based on the (modified) Auger parameter, as originally proposed by Wagner \cite{Wagner1975}. The modified Auger parameter, $\alpha$, of a given cation or anion in an amorphous oxide is defined as ~\cite{Wagner1975, Wagner1988, Moretti1998}:
 
\begin{align}
\label{eq:AP}
    \alpha = E_{\mathrm{b}}^{\rm PE}+E_{\mathrm{k}}^{\rm AE}, 
\end{align}

\noindent where $E_{\mathrm{b}}^\mathrm{PE}$ and $E_{\mathrm{k}}^{\rm AE}$ denote the binding energy of a core-level photoelectron line and the kinetic energy of the corresponding core-core-core Auger line for a given element in the oxide, respectively\footnote{In the original works by Wagner \cite{Wagner1988}, the Auger parameter, $\alpha$, was defined as the difference between the \textit{kinetic} energies of the emitted photo- and Auger electrons only, making it dependent on the incident photon source. This definition was later abandoned, substituted by the \textit{modified} Auger parameter defined in Eq. (\ref{eq:AP}), originally denoted as $\alpha^\prime$, which is now the standard and referred to as $\alpha$.} The Auger parameter thus refers to an energy \textit{difference} between a photoelectron and Auger line, which is independent of the photon source, the defined reference state (e.g., the Fermi level), and any possible static charging during the XPS analysis, making $\alpha$ a robust descriptor of local electronic structure.

Modern photoelectron emission theory requires understanding of initial- and final-state effects during core-ionization, which have recently been linked to the analysis of partial charges on the atoms in the NNCS, as commonly derived from \textit{ab initio} calculations based on DFT \cite{Beck2020}. The partial charges on the central core-ionized atom and its neighbouring atoms before the photoemission event can be quantified using Bader charge analysis. In the initial state (i.e., before the photoemission event), the calculated Bader charges enable the accurate determination of the Madelung potentials at individual sites within the crystal lattice. In the final-state (i.e, upon core-hole creation by photoemission), the calculated Bader charges can be used to predict electronic screening of the localized core-hole by atomic and extra-atomic relaxation, further denoted as $R^\text{a}$ and $R^\text{ea}$, respectively. The atomic relaxation energy, $R^\text{a}$, includes screening contributions of both the core and valence electrons of the core-ionized atom \cite{Moretti1998}. The core-electron contribution to $R^\text{a}$ can typically be assumed to be independent of the chemical state. The valence-electron contribution to $R^\text{a}$ can be considered constant, provided that the number of valence electrons of the core-ionized atom is preserved in its final state. If valence charge adjustment of the core-ionized atom is the dominant contributor to the core-hole screening mechanism, simple linear relations between the Auger parameters of a given cation in the compound and its effective $q_{\text{B}}$ in the ground state can be established \cite{Moretti2020}. In this regard, it is noted that the valence charge of a core-ionized atom in a solid will differ from that of the same isolated atom in a vacuum due to the specifics of chemical bonding with its nearest-neighbouring ligands (i.e., due to hybridization of the valence shells involved in chemical bonding). 

The \textit{extra-atomic} relaxation energy, $R^\text{ea}$, includes screening contributions of the nearest-neighbour ligands around the core-ionized atom \cite{Moretti2020}. Hence, $R^\text{ea}$ reflects non-local screening of the localized core-hole through polarization of nearest-neighbor ligands, which strongly depends on the local structural and chemical environment in its NNCS. The amount of final-state relaxation provided through these mechanisms strongly depends on the amount of valence charge on the core-ionized atom and on the electronic polarizability of the nearest-neighbor ligands, as schematically illustrated in Fig. \ref{fig:Polarizability}(b). The corresponding Auger parameter shift, $\Delta\alpha$, between two different local chemical states of the core-ionized atom in the solid can be estimated from the respective difference in the core-hole relaxation energy, $\Delta R$:
\begin{align}
\label{eq:APshift}
    \Delta\alpha= 2\Delta R=2(\Delta R^\text{a} +\Delta R^\text{ea}), 
\end{align} 
As discussed above, $\Delta R^\text{a}$ can be neglected if the valence charge of the core-ionized atom remains constant between its initial ground state and its final state. Hence, although both initial and final state relaxation mechanisms contribute to the absolute value of the Auger parameter, the Auger parameter shift between two different chemical states is generally only defined by final-state relaxation effects (i.e., provided the number of valence electrons of the core-ionized atom is preserved in its final state \cite{Moretti1998}). For the correct interpretation of Auger parameter shifts of metal cations in solid compounds, it is essential to distinguish the cases dominated by atomic versus extra-atomic relaxations. While atomic relaxation requires advanced quantum mechanical treatment, extra-atomic relaxation can be well described by classical electrostatic models \cite{Moretti1998,Beck2020}.

\subsection{Application to Al oxides}

A \textit{full} chemical state analysis of both the cations and anions in aluminum oxides at \textit{near-constant probing depth} is typically performed by measuring the Al 2p photoelectron and the Al KL$_{23}$L$_{23}$ Auger lines to determine the Al Auger parameter ($\alpha_{\rm Al} = E_{\mathrm{b}}^{\rm Al 2p}+E_{\mathrm{k}}^{\rm Al KL_{23}L_{23}}$), and the O 1s and the O KL$_{23}$L$_{23}$ lines to assess the O Auger parameter ($\alpha_{\rm O} = E_{\mathrm{b}}^{\rm O 1s}+E_{\mathrm{k}}^{\rm O KL_{23}L_{23}}$) \cite{Jeurgens2002, Cancellieri2024}. 
However, the use of oxygen Auger parameters is debated in the literature \cite{Moretti1992, Matthew1997, Moretti1998}, because the O 2p (and, to a lesser extent, the O 2s) states involved in the Auger emission process are (partially) delocalized \textit{valence} levels rather than localized \textit{core levels}, which affects the charge relaxation mechanisms. Indeed, some previous experimental studies have convincingly demonstrated that, despite being difficult to interpret, oxygen Auger parameters are highly sensitive and effective in tracing the development of long-range order during crystallization of amorphous oxides at elevated temperatures ($T \geq$ 200 $^{\circ}$C) \cite{Snijders2005, Jeurgens2008pol, bakradze2011}. By contrast, substantial incorporation of H into ALD amorphous alumina films grown at temperatures up to 200 $^{\circ}$C (i.e., up to the onset temperature for crystallization \cite{Snijders2005, Reichel2008a}), only resulted in negligible O Auger parameter shifts within experimental uncertainty, despite significant changes in composition and density of such oxides were shown to take place \cite{Cancellieri2024}. Unfortunately, a unified theoretical assessment of oxygen Auger parameters is still missing in the literature. Hence, all further discussions and results of this work are fully dedicated to Al Auger parameter shifts only.

We start the theoretical assessment of Al Auger parameters by considering the contribution of the valence charge of the core-ionized Al atom on the atomic relaxation energy $R^\text{a}$. While this has not been done in the past, all the necessary ingredients are present in the literature. A reference Auger parameter of 1454.0 eV was derived by Moretti for Al$^{3+}$ gas, based on the relation of Auger parameter to the refractive index of a compound \cite{Moretti1998}. Such a theoretical reference is quite handy as both atomic and extra-atomic relaxation energies that are enabled by valence electrons of the core-ionized atom and its neighboring ligands are essentially zero. This case, however, is different from the Al$^0$ free atom gas, where only the extra-atomic relaxation contribution is missing, while atomic relaxation is still enabled by three valence electrons. The Auger parameter of the latter has not been reported in the literature, but Al 2p photoelectron \cite{Huttula2009} and KLL Auger spectra \cite{Jänkälä_2007} have been measured separately, with corresponding peak maxima of 81.46 eV (on the binding energy scale) and 1373.36 eV (on the kinetic energy scale), respectively. The resulting Al$^0$ Auger parameter for Al$^0$ free atom gas equals 1454.82 eV, which is shifted +0.82 eV with respect to Moretti's Al$^{3+}$ reference. It follows that atomic relaxation by valence electrons on (isolated) core-ionized Al atom can induce an Auger parameter shift as large as 0.27 eV per valence electron. Al-oxides, Al-hydroxides, and Al-nitrides are characterized by Al Auger parameter values, $\alpha_{\rm Al} = E_{\mathrm{b}}^{\rm 2p}+E_{\mathrm{k}}^{\rm KL_{\rm 23}L_{\rm 23}}$, in the range from 1460 to 1462.7 eV \cite{Wagner1982,Kameshima2000,Cancellieri2024,Pshyk2025}, while the Auger parameter value for Al metal of about 1466 eV is even higher  \cite{Wagner1982,Powell2012}. It is thus safe to assume that much larger Auger parameter shifts for Al-oxide, hydroxides, and nitrides in the range from 6 to 12 eV with respect to Al$^{3+}$ gas are dominated by extra-atomic relaxation, i.e., $R^\text{ea} \gg R^\text{a}$. We can prove that the contribution of atomic relaxation is truly negligible in semiconductor materials by considering the results of Bader charge analysis, indicating that Al has a Bader charge of 2.47 in Al oxides and hydroxides \cite{Gramatte2025} and 2.39 in AlN \cite{Chaumeton2016}. This means that only around half an electron charge is residing at the valence shells of the Al ions in such compounds, enabling only around 0.15 eV of Auger parameter shift through atomic relaxation. Thus, for Al compounds, Eq.(2) reduces to \begin{align}
\label{eq:Ap_shift}
    \Delta \alpha\ = 2\Delta R^{\mathrm{ea}}.
\end{align}
In this limiting case, any measured shift, $\Delta\alpha_{\rm Al}$, between two different amorphous alumina polymorphs (e.g., with low and high H content) is indicative of tiny differences in the NNCS of the core-ionized Al atoms (as defined by bond lengths, bond angles, and coordination number) and/or a change in the (averaged) ligand polarizability; the corresponding difference in extra-atomic relaxation energy can be estimated using relatively simple electrostatic models \cite{Moretti1998, Ambrosio2017}.

\subsection{Electrostatic models of extra-atomic relaxation}
\label{sub:TheoryElectrostaticModels}

The application of electrostatic models requires the Auger parameter shift to be dominated by extra-atomic relaxation, i.e., $R^\text{ea} \gg R^\text{a}$ (\textit{Assumption \#1}), which is the case for Al as discussed above. However, this assumption does not work for many transition metals \cite{Moretti1998,Moretti2020} and should be carefully evaluated before using electrostatic models to interpret Auger parameter shifts. 
Historically, researchers have proposed numerous empirical correlations to link the Auger parameter shift to the electronic polarizability of the chemical environment surrounding the core-ionized atom \cite{Fiermans1975, Kohiki1987, Edgell1990, Jeurgens2008pol}. 
Moretti \cite{Moretti1990} developed a simple electrostatic model to calculate the extra-atomic polarization energy, $R^{\text{ea}}$, of the first-neighbor ligand shell around core-ionized cations and thereby the resulting Auger parameter shift $\Delta\alpha = 2R^{\text{ea}}$. In the proposed model, the final-state polarization process is characterized by classical electrostatic calculations. These calculations aim to determine the total electric field experienced by the ligands upon the creation of a core hole in a central atom. This total field includes contributions from both the central positive charge of the core-ionized atom and the electric fields resulting from induced dipoles on the ligands within the closest-neighbor shell \cite{Moretti1998} (\textit{Assumption \#2}). 
Although a strong approximation resulted from the analysis of molecular systems, the nearest-neighbors models have been shown to work equally well for crystalline and amorphous solids \cite{Moretti1998}. The interactions between the core-ionized atom and its nearest neighbors are encapsulated by the electronic polarizability volumes of the ligands (in \AA$^3$), denoted as \(\zeta_{i}\). The parameter \(\zeta_{i}\) serves to quantify the magnitude of the dipole moment 
\begin{align}
    M_{\text{i}}=4\pi\varepsilon_0\zeta_{i}F_{\text{L}_{i}}
\end{align}
induced within the $i \rm ^{th}$ ligand as a direct response to the electric field, $F_{L_{i}}$, generated by the core-ionized atom (see Fig. \ref{fig:Polarizability}c), with $\varepsilon_0$ being the dielectric constant. This selective focus on the first coordination shell and the corresponding electronic polarizabilities provides a nuanced understanding of the electrostatic interactions facilitated by the presence of a core hole, highlighting the model's utility in studying the electronic structure and bonding properties of materials.

\subsubsection{Complete model formulation}\label{subsub:full_model_formulation}
 
Moretti \cite{Moretti1998} developed a comprehensive electrostatic model that utilizes a series of geometric assumptions to predict the extra-atomic relaxation energy. The model is concisely captured by the equation:
\begin{align}
\label{eq:full_model}
    \Delta \alpha  = 2R^{\text{ea}} = 14.4 \sum_{i} \zeta_i/r_i^2 \sum_j C_{ij}/r_j^2,
\end{align}
where \(\zeta_i\) symbolizes the electronic polarizability volume of the \(i^{th}\) ligand, and 14.4 is the scaling factor to get the eV units. In this expression, \(r_i\) and \(r_j\) refer to the distances from the core-ionized atom to the \(i^{th}\) and \(j^{th}\) ligand, respectively. The spatial distribution of these distances and related parameters is depicted in Fig. \ref{fig:Polarizability}(c).
$C_{ij}$ contains all further geometric information to calculate $\Delta \alpha $ and is given by \cite{Moretti1998}
\begin{align}
    \textit{\textbf{C}}=\textit{\textbf{B}}^{-1},
\end{align}
with 
\begin{align}
    B_{ij}=\begin{cases}
    1, & \text{if $i=j$}.\\
    \zeta_j T_{ij}/(r_i \cos \Tilde{\theta}_{ij}+r_j\cos\theta_{ij})^3, & \text{otherwise}.
  \end{cases}
\end{align} 
and $T_{ij}$ given as 
\begin{align}
    T_{ij}=\cos \theta_{ij} \sin \Tilde{\theta}_{ij} \sin{(\Tilde{\theta}_{ij}+\theta_{ij})}+(3\cos^2\theta_{ij}-1)\cos{(\Tilde{\theta}_{ij}+\theta_{ij})}.
\end{align}
Within the framework of the aforementioned equations, the terms \(\theta_{ij}\) and \(\Tilde{\theta}_{ij}\) play a pivotal role in quantifying the angular relationship between ligands and the core-ionized atom, as visualized in Fig. \ref{fig:Polarizability}(c). Specifically, these parameters capture the spatial orientation of the ligand relative to another ligand and the core-ionized atom, thereby reflecting the geometric nuances of their interactions. Consequently, \(T_{ij}\) emerges as a crucial term designed to encapsulate the local environmental characteristics of an atom, employing geometric functions to provide a comprehensive description. This approach enables a detailed assessment of how the spatial arrangement and angular disposition of ligands affect the electrostatic potential and, by extension, the electrostatic properties of the system under study.

\subsubsection{Simplified model formulation}
\label{subsub:simplifiedModelFormulation}

Building upon the comprehensive framework for predicting \(\Delta R^{\text{ea}}\) with the use of \(T_{ij}\), Moretti also introduced a simplified version of the model. This streamlined approach, detailed in his seminal works \cite{Moretti1990, Moretti1991, Moretti1998}, relies on specific assumptions relevant to ordered materials that exhibit high symmetry. Central to this simplification is the postulate that the angular relations between the ligands and the core-ionized atom are uniform, denoted as \(\theta_{ij} = \Tilde{\theta}_{ij}\), and that the electronic polarizabilities of all ligands are equal, \(\zeta_i = \zeta_j = \zeta\). Additionally, it assumes that the distances from the core-ionized atom to all ligands are identical, represented by \(r_i = r_j = r\). Under these conditions, the model yields a more accessible expression for estimating the extra atomic relaxation energy, encapsulating the effects of symmetry and uniformity in periodic crystal lattices on its electrostatic properties. The simplified model for the Auger parameter shift with respect to the gas phase is defined as:

\begin{align}
\label{eq:AP_simpliefied}
    \Delta\alpha=2 R^{\text{ea}}=\frac{14.4n\zeta}{rD\zeta+r^4},
\end{align}

\noindent where the factor

\begin{align}
    D=\sum_{j\neq i}\frac{1+\cos^2\theta_{ij}}{8\cos^3\theta_{ij}}
\end{align}

\noindent occupies a central role in the computation of Auger parameter shifts, primarily because it accounts for the dipole-dipole repulsion among the induced dipoles on the ligands.

In the context of atomistic simulations targeting non-symmetric solids with variable coordination environments, as for the amorphous alumina polymorphs in the present study, we adopt the methodology wherein the relationship delineated in Eq. (\ref{eq:AP_simpliefied}) is presumed to be applicable (\textit{Assumption \#3}). This applicability is facilitated through the employment of time- and structure-averaged values for all pertinent geometric properties derived from the simulation's production runs.  

For each sample with different H content, we account for the different fractions of the O$^{2-}$ and OH$^-$ nearest-neighbor ligands with polarizabilities $\zeta_{\mathrm{O}}$ and $\zeta_{\mathrm{OH}}$, respectively, by assuming that the average polarizability simply follows the rule of mixtures (\textit{Assumption \#4}): 

\begin{align}
\label{eq:averages_polarizabilities}
    \zeta=\zeta_{\mathrm{O}} (1-f_{\mathrm{OH}})+\zeta_{\mathrm{OH}} f_{\mathrm{OH}},
\end{align}

\noindent where $f_{\mathrm{OH}}$ is the time-averaged fraction of OH$^-$ nearest-neighbor ligands \cite{Gramatte2025}.
Next, we aim to validate all 4 assumptions listed above using numerical methods with atomic layer deposited amorphous alumina as a model system. It will be convincingly demonstrated how such a framework can enable predictive modeling of extra-atomic relaxation energy and Auger parameter chemical shifts as a function of temperature.


\section{Numerical methods and computational details}

\begin{figure}[H]
    \centering
    \includegraphics[width=\textwidth]{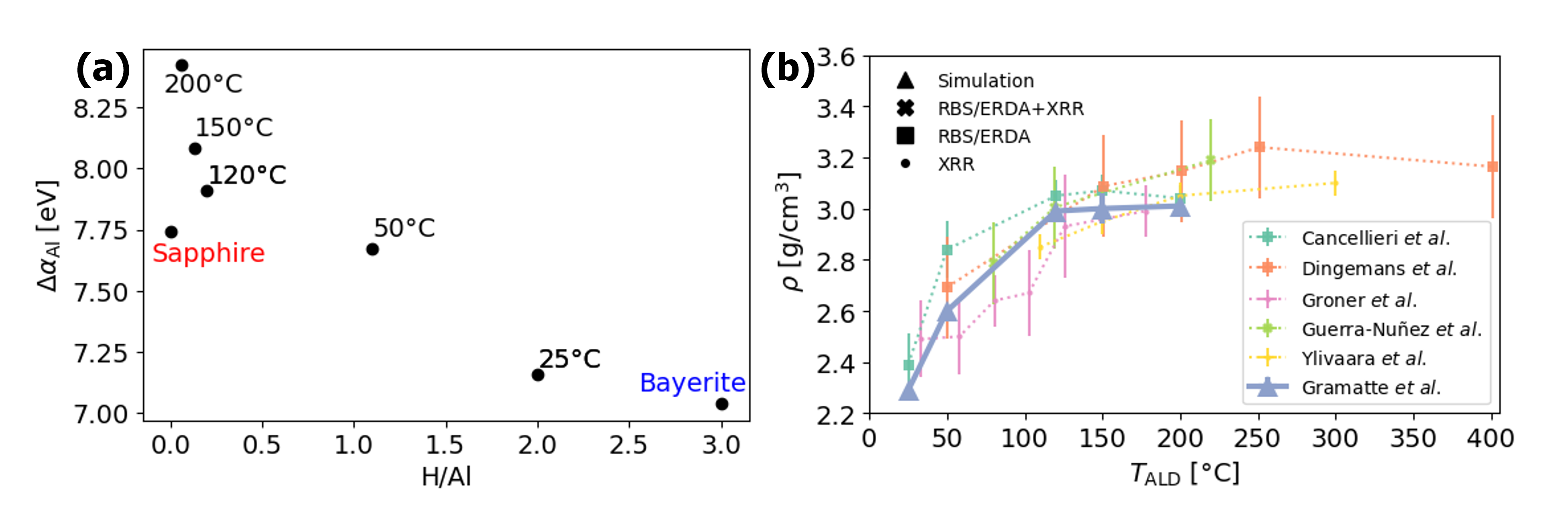}
    \caption{(a) $\Delta\alpha_{\rm Al}$ as a function of the stoichiometry H/Al. (b) Density $\rho$ as a function of the deposition temperature, $T_{\mathrm{ALD}}$, from various publications using thermal ALD, with different methods for density determination: XRR as crosses, RBS/ERDA as squares, and a combination of RBS/ERDA+XRR as diamonds~\cite{Cancellieri2024,Dingemans2010,Groner2004,Guerra-Nunez2017,Ylivaara2014}. The simulated densities used in this work, previously validated in Ref.~\cite{Gramatte2025}, are highlighted as triangles above all experimental values.}
    \label{fig:density_comp_andshift}
\end{figure}

In this work, we focus specifically on the dataset provided by Cancellieri et al. \cite{Cancellieri2024}, which reports systematic measurements of Auger parameter shifts and H/Al stoichiometry of amorphous alumina as a function of the ALD process temperature $T_{\mathrm{ALD}}$, as shown in Fig. \ref{fig:density_comp_andshift}(a). Further details on the experimental methodology and data can be found in the Supplementary Material. The stoichiometries and deposition temperatures extracted from that study, summarized in Table S1 of the SM, serve as the foundation for our atomistic simulation approach. Following the methodology introduced in Ref.~\cite{Gramatte2025}, we constructed ALD alumina polymorphs with varying hydrogen content, based on the H/Al ratios determined experimentally. This approach yields densities that closely reproduce the experimental trends across a wide ALD growth temperature range, as highlighted with blue triangles in Fig.~\ref{fig:density_comp_andshift}(b).

\subsection{Molecular dynamics simulations}
\label{Sec:atomistic_Simulations}

In our study, the methodology used was centered on molecular dynamics (MD) simulations, a technique that is pivotal to probing the atomic-scale interactions and structural nuances of materials. MD simulations were carried out using Large-scale Atomic/Molecular Massively Parallel Simulator (LAMMPS) \cite{Thompson2022}. In absolute zero calculations, the conjugate gradient method was used to obtain fully relaxed samples. At higher temperatures, the Nose-Hoover thermostat and the Parinello-Rahman barostats were used for temperature and pressure control with damping coefficients of 0.05 ps and 0.5 ps, respectively. The lower-than-default timestep was set to 0.5 fs, to avoid artifacts related to unrealistic hydrogen mobility in amorphous structures and to ensure proper equipartition between different degrees of freedom in the system \cite{Asthagiri2024}. Periodic boundary conditions were applied along all dimensions to model bulk phases, excluding surface effects. Following the simulation phase, structural processing and analysis were conducted using pymatgen \cite{Jain2011}, ASE \cite{Larsen2017}, and OViTo \cite{Stukowski2010b}, as similarly described in the SI of \cite{Gramatte2025}.

For MD simulations, a universal graph neural network potential (NNP), specifically Matlantis's PreFerred Potential (PFP) \cite{Takamoto2022}, was used. We recently demonstrated the applicability of this type of graph NNPs for the accurate modeling of amorphous alumina \cite{Gramatte2024}, and will further validate in the SI Fig. 3 that PFP shows superior accuracy for defective, porous, and amorphous alumina.  PFP stands out because of its ability to simulate a broad spectrum of molecular and crystalline systems, extending its utility to uncharted materials. It supports simulations that involve combinations of up to 72 elements, supported by a comprehensive training dataset comprising more than 32 million structures consistently derived from high-quality DFT calculations \cite{Matlantis2022}. As with any other potential, its applicability to the system and property of interest is first tested and justified, as discussed in \cite{Gramatte2024b,Gramatte2025}. After the inclusion of Van der Waals dispersion correction, namely the Becke and Johnson (BJ) D3 correction \cite{Grimme2010,Grimme2011}, the potential accurately reproduces the structures and properties of reference crystalline phases and liquid alumina. Although the PFP model does not impose formal restrictions on stoichiometry, we impose charge neutrality to improve reliability. This may cause slight deviations from experimental ALD alumina stoichiometries but ensures more accurate results.

To generate realistic models of amorphous oxide polymorphs with varying hydrogen contents, we employed the simulation approach previously developed and validated in Ref. \cite{Gramatte2025}. This method enables direct control over hydrogen incorporation while preserving agreement with experimental densities. 
In brief, the initial structures were derived from the crystalline Al-trihydroxide phase, bayerite. Hydroxyl groups and their associated hydrogen atoms were randomly removed in stoichiometric OH-H to maintain overall charge neutrality. The number of removed pairs was adjusted to match the target H/Al ratios determined experimentally.
To promote structural relaxation and amorphization, the lattice vectors of the resulting defective structures were isotropically scaled, leading to densification and bond reconstruction during equilibration.

The systems were equilibrated in the NVT ensemble at the deposition temperature (T$_\mathrm{ALD}$) for 50 ps, allowing the defective bayerite structures to transition into amorphous phases. This was followed by a quench to the equilibration temperature ($T_\mathrm{E}$ = 27°C) and a subsequent 50-ps equilibration in the NpT ensemble, during which all six cell degrees of freedom were allowed to relax. An additional 50 ps of equilibration at $T_\mathrm{E}$ in the same ensemble was carried out to determine the time-averaged cell parameters, which were then fixed for the final NVT production runs.
All input files, scripts, and trajectories are available on the Materials Cloud~\cite{Talirz2020} Archive in Ref.~\cite{Gramatte_Auger_data}.  

The resulting densities are reported in Table S1 in the SM, demonstrating excellent agreement between the densities predicted from simulations and measured experimentally, highlighting the reliability of PFP and the new simulation method. 
The same equilibration and production protocol was applied to the crystalline reference structures at $T_\mathrm{E}$ = 27 °C to ensure consistency across amorphous and crystalline systems. During the final production runs, 1000 snapshots were extracted at regular intervals for each structure to enable time-averaged analysis that incorporates thermal fluctuations.

\begin{figure}
    \centering
\includegraphics[width=1\linewidth]{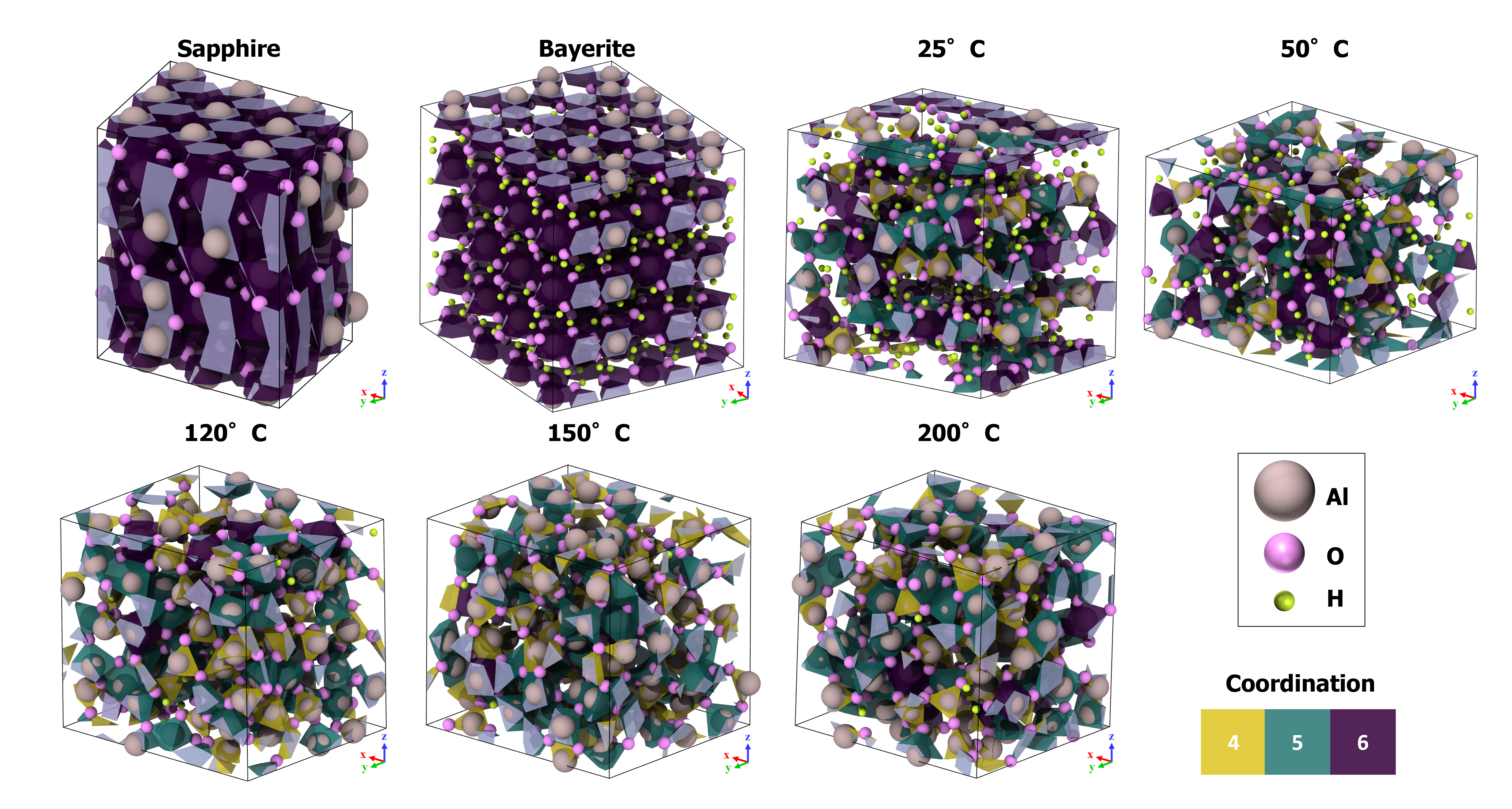}
    \caption{Al-coordination analysis of five amorphous alumina polymorphs with different H/Al ratios and densities as well as of crystalline sapphire and bayerite. Details on simulation cell dimensions and atom counts appear in the SI of Ref.~\cite{Gramatte2025}.}
    \label{fig:CoordinationPolyhedra}
\end{figure}

An overview of the used structures and their Al-coordination environment is given in Fig. \ref{fig:CoordinationPolyhedra}.
Fig. \ref{fig:CoordinationPolyhedra} shows the coordination environment of Al atoms based on their first-neighbor oxygen shells. In the crystalline reference phase Al$_2$O$_3$ (sapphire), Al$^{3+}$ cations are exclusively coordinated octahedrally by O$^{2-}$ anions. In the trihydroxide reference phase -Al(OH)$_3$ (bayerite), Al$^{3+}$ occupies octahedral sites between two stacked planes of close-packed OH$^{-}$ groups.

The amorphous structures generated at a deposition temperature of 27$^\circ$C display a broader distribution of local environments, with nearly equal populations of 6-fold (violet), 5-fold (green), and 4-fold (yellow) [AlO$_n$] polyhedra. This distribution agrees well with experimental observations \cite{dupree1985, Poe1992, Ansell1997, Cui2018, Gutierrez2000, Landron2001a}, confirming that the defective bayerite-derived structures undergo a transition to the amorphous phase upon equilibration at the ALD processing temperature.

Comparative analysis of amorphous polymorphs produced at ALD temperatures of 50$^\circ$C, 120$^\circ$C, 150$^\circ$C, and 200$^\circ$C (all quenched to $T_\mathrm{E}$ = 27$^\circ$C) reveals a systematic trend: increasing the processing temperature reduces the hydrogen content and the proportion of 6-fold coordinated [AlO$_n$] units. At elevated temperatures, 5-fold coordination becomes dominant over 4-fold, indicating a progressive shift in the short-range order around the Al cations with decreasing H/Al ratio (i.e., with decreasing number of OH hydroxyl ligands in the Al NNCS).

\subsection[Delta Kohn-Sham calculations]{XPS calculations using the $\Delta$Kohn-Sham DFT method}
\label{subsec:deltaKS}

Within the $\Delta$KS DFT method~\cite{Cavigliasso1999}, the binding energy $E_\mathrm{B}^{\Delta\mathrm{KS}}$ of a core level $j$ for an atom $i$ is computed as the total energy difference between the excited state $E^*$, where one electron is removed from the core level, and the ground state $E^0$:
\begin{equation}\label{BE}
E_{\mathrm{B},ij}^{\Delta\mathrm{KS}} = E^{*}(n_{c,j}-1,n_v+x)-E^{0}(n_{c,j},n_v)+
\Delta E_{\mathrm{corr}},
\end{equation}
where $n_{c,j}$ and $n_v$ are the populations of the core state $j$ and valence state, respectively; the value of $x$ distinguishes between FCH (Full Core Hole, $x = 0$) \cite{Rößler2003,Hetényi2004}  and XCH (eXcited electron and Core Hole, $x = 1$) \cite{Prendergast2006} treatments. 
The corrective term $\Delta E_{\mathrm{corr}}$ provides absolute $E_{\mathrm{B}}$ comparable to experiments, accounting for both relaxation effects of the remaining core electrons in the presence of a core-hole~\cite{mizoguchi2009}, as well as the approximation related to the usage of semilocal DFT~\cite{Walter2016}. The procedure to compute $\Delta E_{\mathrm{corr}}$ is described in detail in Ref.~\cite{Walter2016} and exemplified for Al in Supporting Information.

%
Unless interested in core-level shifts only, the reference energy level should be taken into account to correctly determine $E_{\mathrm{B}}$ from \autoref{BE}. Within the FCH treatment, $E_{\mathrm{B}}$ is referenced to the vacuum level; using the XCH approximation, the reference is placed at the Fermi level ($\varepsilon_F$) in metals or at the bottom of the conduction band ($\varepsilon_{\text{CB}_{\mathrm{min}}}$) in insulators. In the case of non-metals, $\varepsilon_F$ lies somewhere in the gap, hindering the energy referencing procedure. 
To address this issue, $\varepsilon_F$ in insulators is often approximated as the midpoint of the DFT gap for the ground state system \cite{Walter2016}, 
%
%
leading to the corrected binding energy:
\begin{equation}\label{BE_corr}
E_{\mathrm{B},ij}^{F} = E^1_{\mathrm{B},ij}-\varepsilon^1_{\text{CB}_{\mathrm{min}}}+\varepsilon_F,
\end{equation}
where $E^1_{\mathrm{B},ij}$ (following \autoref{BE}) and $\varepsilon^1_{\text{CB}_{\mathrm{min}}}$ are computed within the XCH. In metals, where $\varepsilon_{\text{CB}_{min}}=\varepsilon_F$, this correction has no effect. In insulators, the correction introduces uncertainty due to the well-known underestimation of the band gap in semilocal DFT. Therefore, \autoref{BE_corr} should be used cautiously in large-gap systems. 

\begin{figure}[H]
    \centering
    \includegraphics[width=1\linewidth]{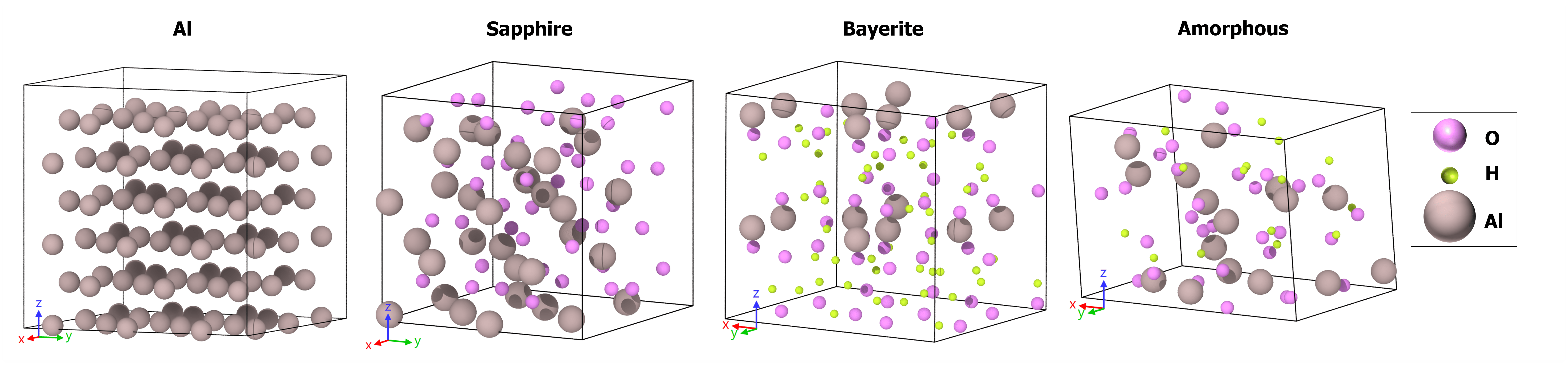}
\caption{
Initial structures used for the $\Delta$KS calculations: crystalline aluminum, $\alpha$-Al$_2$O$_3$ (sapphire), $\alpha$-Al(OH)$_3$ (bayerite), and an amorphous Al$_2$O$_3$ model with H/Al = 1. In both aluminum and bayerite, all Al atoms are symmetry-equivalent; hence, a single $\Delta$KS calculation was performed per structure to determine the binding energy $E_{\mathrm{B}}^{\Delta\mathrm{KS}}$. Sapphire contains two distinct Al sites, so $E_{\mathrm{B}}^{\Delta\mathrm{KS}}$ was computed as the average of two $\Delta$KS calculations. In the amorphous model, all Al atoms are structurally unique; therefore, a separate $\Delta$KS calculation was conducted for each site, and $E_{\mathrm{B}}^{\Delta\mathrm{KS}}$ was obtained by averaging over all results.
}

    \label{fig:structures_dft}
\end{figure}

To facilitate and streamline $\Delta$KS-based XPS calculations, we used the automated AiiDA workflow available through the \texttt{aiida-qe-xspec} package~\cite{aiida_qe_xspec}, integrated tightly with the \textsc{Quantum ESPRESSO} package~\cite{Giannozzi:2009, Giannozzi:2017, Giannozzi:2020} for DFT simulations. The workflow is accessible via the AiiDAlab\cite{Yakutovich2021} Quantum ESPRESSO app graphical interface \cite{aiidalab_qe} and fully automates the setup and execution of total energy calculations needed for core-level binding energy estimation. 
For our simulations, we used the default balanced protocol settings of the AiiDAlab Quantum ESPRESSO app, as defined in Ref. ~\cite{nascimento2025}. The exchange–correlation functional was treated with the PBE approximation\cite{Perdew1996}. Pseudopotentials taken from the SSSP PBE efficiency v1.3 library~\cite{prandini_precision_2018, prandini_2023_rcyfm-68h65} were used to describe the ground-state properties, whereas core-hole pseudopotentials for Al (1s, 2s, and 2p) and O (1s) were generated on purpose to describe core-ionized atoms. 
The calculation of $E_\mathrm{B}^{\Delta\mathrm{KS}}$ was carried out by adopting the XCH treatment (i.e., the removed core-electron is placed in the valence), which shows to converge faster with the supercell size. For these calculations, spin polarization was enabled. 
Instead, FCH treatment was adopted for analyzing the charge density redistribution. This corresponds to the complete removal of one electron from the system, which is expected to better reflect the localized electronic relaxation around the core hole.
For the FCH calculations, kinetic energy cutoffs of 50 Ry ($\sim 680$ eV) for the wavefunctions and 400 Ry ($\sim 5442$ eV) for the charge density were used to define the plane-wave basis. Electronic occupations were treated via Gaussian smearing with a width of 0.05 Ry ($\sim 0.68$ eV). 
Brillouin‐zone integrations used a Monkhorst–Pack grid of $10 \times 10 \times 10$ k‐points for bulk sapphire, corresponding to a k-point linear spacing of approximately 0.036$~\text{\AA}^{-1}$. This served as a reference to maintain a consistent k‐point sampling density across supercells of varying size. 
To reduce computational cost, the $\Delta$KS calculations were carried out using smaller simulation cells than those introduced in Sec.\ref{Sec:atomistic_Simulations}. The $\Delta$KS structures, shown in Figure \ref{fig:structures_dft}, were either generated following the same procedure described previously \cite{Gramatte2025}, or taken from the same literature sources but based on reduced initial cell sizes.

\subsection{Bayesian optimization of ligand polarizabilities}
\label{subsub:Bayesian_methods}

Bayesian optimization is a powerful technique for optimizing objective functions that are expensive to evaluate. It operates by constructing a posterior distribution of functions, typically using a Gaussian process, which serves as a probabilistic model for predicting the objective function's outputs based on its inputs. As observations accumulate, this posterior distribution is refined, enhancing the algorithm's capacity to identify promising regions within the parameter space for further exploration.
For a comprehensive understanding of this method, readers are referred to the relevant literature \cite{Seeger2004,Brochu2010,Rasmussusen2005}.

By applying the analysis based on Eq. (\ref{eq:averages_polarizabilities}) or Eq. (\ref{eq:AP_simpliefied}) to 1000 snapshots from the MD production run for each sample, we can compare the predicted time- and structure-averaged Al Auger parameter shifts, \(\Delta\alpha_{\text{Al,pred}}\), with our experimentally-measured data, \(\Delta\alpha_{\text{Al,exp}}\), as presented in the SI Table 1. This formulation enables the establishment of an acquisition function, \(F\), which solely depends on \(\zeta_{\text{OH}}\) and \(\zeta_{\text{O}}\), expressed as:
\begin{equation}
\label{f_function}
    F = -\sum_{\text{samples}} \left(\Delta\alpha_{\text{Al,exp}} - \Delta\alpha_{\text{Al,pred}}(\zeta_{\text{OH}},\zeta_{\text{O}})\right)^2.
\end{equation}
This function $F$ is then maximized using Bayesian optimization, constrained within the experimentally reported range of polarizability volumes for metal oxides (0.9--3.2~\AA$^3$), which are assumed to be equal for both ligands \cite{Tessman1949}.

\section[Results and Discussion]{Results and Discussion}
\subsection{Binding energy predictions using the \texorpdfstring{$\Delta$}{Delta}Kohn-Sham DFT method}

\begin{figure}
    \centering
    \includegraphics[width=0.8\linewidth]{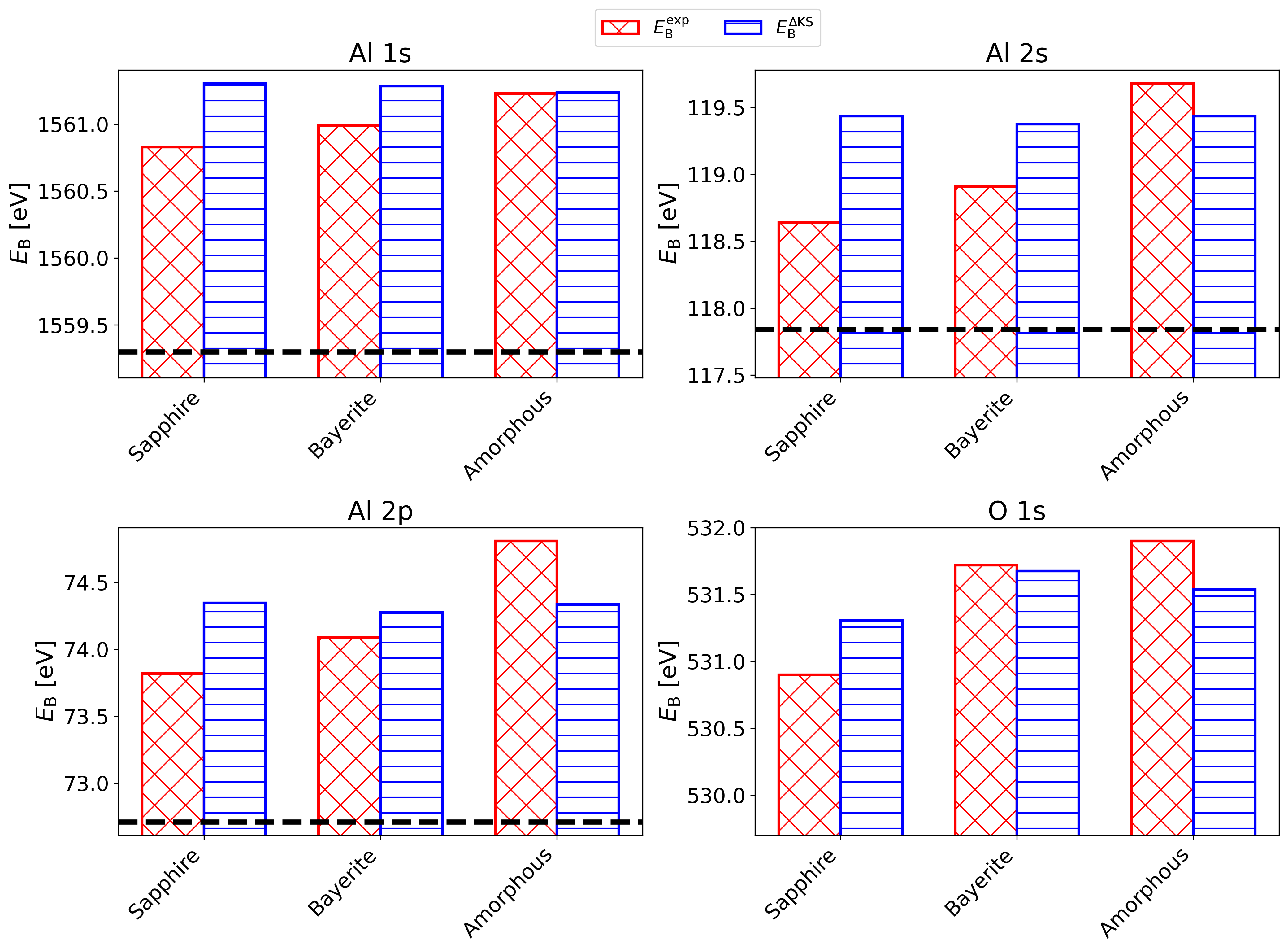}
    \caption{Histograms of core electron binding energies $E_{\mathrm{B}}$ for four materials --Al, sapphire, bayerite, and the amorphous alumina sample with H/Al = 1 (Amorphous)-- are shown with adapted $y$-axes. Experimental values $E_{\mathrm{B}}^{\mathrm{exp}}$ are shown in red, and calculated values using the $\Delta \mathrm{KS}$ and XCH method $E_{\mathrm{B}}^{\Delta \mathrm{KS}}$ are shown in blue. Binding energies measured experimentally:  Al 1s (using Cr X-rays),  Al 2s (using Al X-rays),  Al 2p (using Al X-rays),  O 1s (using Al X-rays). The dashed black line corresponds to $E_{\mathrm{B}}^{\mathrm{exp}}$ of Al metal. }
    \label{fig:Binding_energie_hist}
\end{figure}

In addition to the experimental core electron binding energies $E_\mathrm{B}^{\mathrm{exp}}$ published in~\cite{Cancellieri2024} and summarized in Table S1 in the SM, we computed core level binding energies using the $\Delta \mathrm{KS}$ method described in \autoref{subsec:deltaKS}.
Figure~\ref{fig:Binding_energie_hist} compares the experimental binding energies $E_\mathrm{B}^{\mathrm{exp}}$ with $E_\mathrm{B}^{\Delta\mathrm{KS}}$ across three aluminum-containing materials: sapphire (Al$_2$O$_3$), bayerite (Al(OH)$_3$), and an amorphous sample with H/Al = 1, labeled as Amorphous. The Al metal binding energies were used for performing the fitting as described in the Supporting Information, and the values are shown in the figure as horizontal lines. 
For all the core levels examined (Al 1s, Al 2s, Al 2p, and O 1s), the computed binding energies reproduce the quantitative range but not the qualitative trend observed in the experimental data: binding energies increase progressively from the metallic to the more oxidized, non-metallic environments. This shift is especially pronounced between metallic Al and the oxide phases, and the direction of the shift is correctly predicted across all core levels. However, quantitative discrepancies arise among the non-metallic phases. Specifically, for the Al 2s, Al 2p, and O 1s levels, the calculated $E_\mathrm{B}^{\Delta\mathrm{KS}}$ values for bayerite and the amorphous sample are slightly overestimated relative to sapphire when compared to experiment, resulting in an incorrect ordering of binding energies among the non-metals. These deviations suggest that, while the $\Delta$KS method successfully captures the physics of metal-to-insulator transitions, it requires a proper binding energy alignment to resolve structurally similar insulating phases (i.e., to distinguish between different polymorphic oxide and/or hydroxide phases).

High uncertainty affects both experimental and computational binding energies. In XPS, for example, differential charging of the sample surface can induce spurious binding energy shifts, especially in insulating oxides and amorphous phases \cite{zhang_jeu2025}. Even though dual-beam flood-guns are nowadays employed to compensate for the photoemitted electronic charge, incomplete charge compensation in combination with incorrect energy scale calibration procedures can lead to experimental errors in the absolute binding energies of insulating compounds as large as 2-3 eV \cite{zhang_jeu2025}. Additionally, the experimental determination of the Fermi level reference, which is often calibrated via conductive standards or secondary electron cutoffs, can introduce systematic offsets if not precisely controlled \cite{Baer2020}. On the computational side, uncertainties arise from the choice of the exchange-correlation functional and the treatment of the core hole effect in the $\Delta \mathrm{KS}$ approach. A central challenge in computing absolute binding energies is the definition of the reference energy, particularly in systems with a finite band gap, where the Kohn–Sham Fermi level is ill-defined. A common strategy, which is used in this calculation, is to approximate the Fermi level as the midpoint of the DFT gap in the ground state. However, this midpoint assumption is itself a significant source of error, especially since the band gap is typically underestimated by semi-local functionals such as PBE due to self-interaction errors ~\cite{mori2008localization}. Recognizing these factors clarifies the current accuracy limits and identifies ways to improve experimental calibration and methodology.

\begin{figure}[H]
    \centering
    \includegraphics[width=1\linewidth]{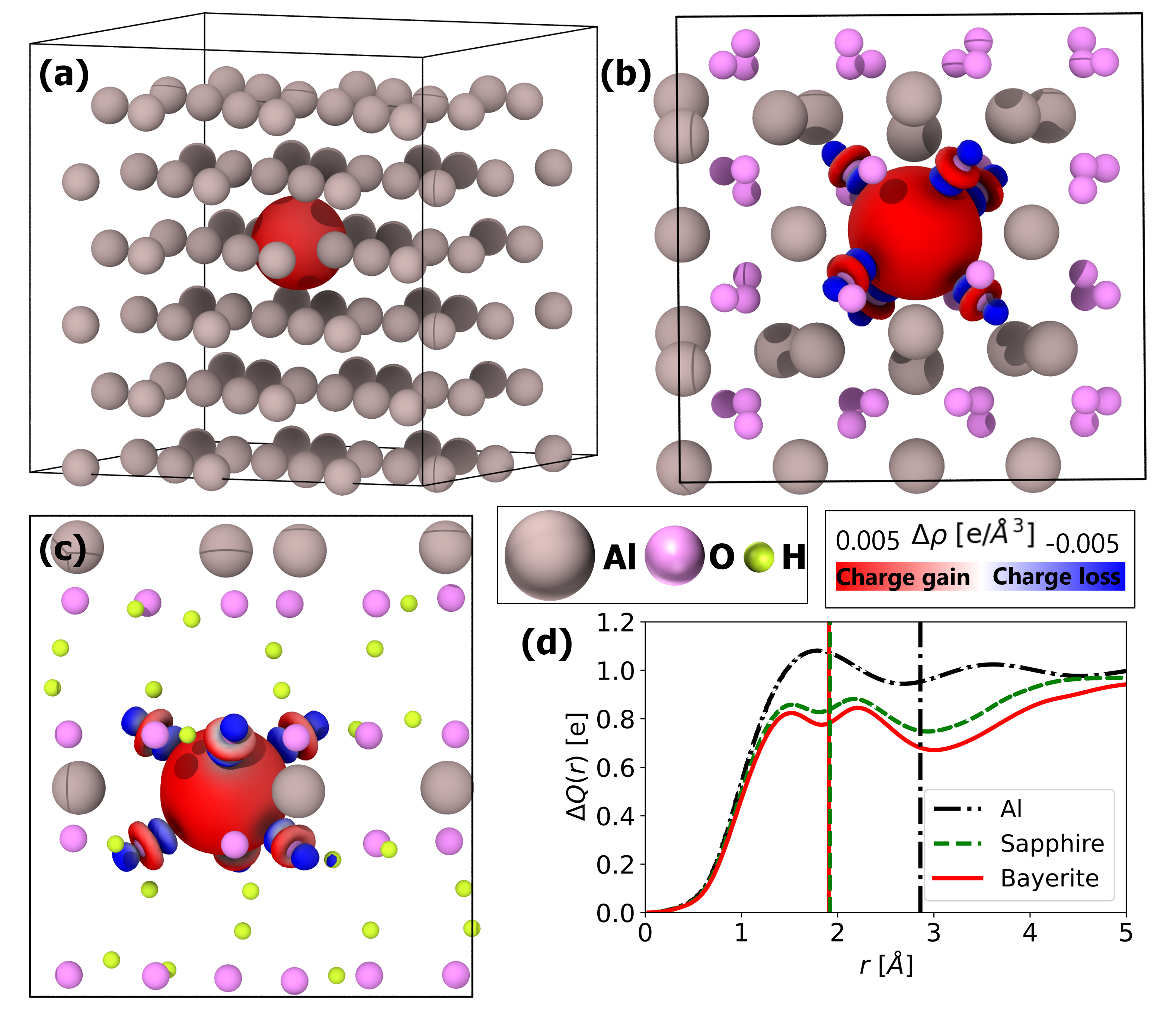}
    \caption{(a) Valence electron density difference $\Delta \rho_{\mathrm{v}}$ for metallic Al between the ground-state DFT calculation and the excited-state calculation, plotted as isosurfaces. Red regions indicate charge accumulation, and blue regions indicate charge depletion, caused by the creation of a core hole in a single Al atom. 
(b) Electron density difference for sapphire, analogous to panel (a).  
(c) Electron density difference for bayerite, analogous to panel (a).
(d) Radial integrated valence charge difference $\Delta Q(r)$ as a function of radial distance $r$ from the center of the core-ionized atom ($r = 0$) for Al (black), sapphire (green), and bayerite (red). The vertical black line visualizes the mean Al-Al nearest-neighbor distance, whereas the green and red lines visualize the mean bond length of the Al-O nearest-neighbor distance, which is very similar in sapphire and bayerite. 
}
    \label{fig:charge_diff_and_cumul}
\end{figure}

To further probe the role of final-state relaxation and screening effects, Figs.~\ref{fig:charge_diff_and_cumul}(a)-(c) present the valence charge density difference $\Delta\rho_{\mathrm{v}}$  between the FCH and ground-state calculations. The red regions indicate charge accumulation, and the blue regions indicate charge depletion around a core-ionized Al atom. In elemental metallic Al, the delocalized electron gas rapidly neutralizes the core hole, yielding a compact, nearly spherical screening cloud. By contrast, the low free-carrier density in sapphire and bayerite produces a more extended and diffuse redistribution of charge around the ionized site. These contrasting behaviors are directly visualized in the electron density difference isosurfaces: the metal shows concentrated, isotropic features, whereas the oxides exhibit broader, ligand-directed regions of accumulation and depletion. Such material-specific screening mechanisms underscore the importance of final-state relaxation processes in accurate core-level spectroscopy. Even more importantly, these results validate Assumption \#2 in the definition of the electrostatic model (Section \ref{sub:TheoryElectrostaticModels}), stating that ligand polarization effects and non-local screening are limited to the first neighbor shell (see Fig.~\ref{fig:charge_diff_and_cumul}(b,c)). 

Fig.~\ref{fig:charge_diff_and_cumul}(d) shows the radial integrated valence charge difference $\Delta Q(r)=4\pi\int_0^r\Delta\rho_{\mathrm{v}}(r^\prime){r^\prime}^2dr^\prime$, where $\Delta\rho_{\mathrm{v}}(r)$ is the valence charge difference defined above as a function of distance $r$ from the core-ionized site.  The integrated charge redistribution, derived from a voxel-based analysis analogous to the Bader charge-partitioning method \cite{Henkelman2006,Sanville2006,Tang2009}, is displayed for metallic Al (black), sapphire (grey), and bayerite (red), and illustrates the differing spatial extents of screening charge in metallic versus oxide environments. The inclusion of a compensating homogeneous jellium background ensures overall charge neutrality of the simulation cell but also perturbs the balance between atomic and extra–atomic relaxation, causing the integrated valence charge difference $\Delta Q(r)$ with respect to the ground state to converge to one electron rather than zero. Upon core–hole creation, this electron localizes near the ionized atom, partially screening the positive charge. As shown in Fig.~\ref{fig:charge_diff_and_cumul}(d), the initial accumulation of compensating charge within the nearest–neighbor shell is similar for metallic Al, sapphire, and bayerite. 
After curves separation, metallic Al continues to attract additional screening charge—exceeding one elemental charge before converging slowly to one after the second–nearest–neighbor radius—reflecting the metallic environment's high free–electron density. In contrast, sapphire reaches a first peak of approximately $0.85|e|$ at approximately 1.7~\AA    and a secondary, slightly lower maximum around 2.3~\AA, before converging to one; bayerite follows a similar profile with marginally reduced amplitude, consistent with its lower density of non-localized electronic charge.

These material–specific screening profiles correlate directly with shifts in the Auger parameter $\Delta\alpha_{\mathrm{Al}}$, as reported in Table S1 in the SM: 12.0~eV for Al, 7.7~eV for sapphire, and 7.0~eV for bayerite. The proportionality between integrated charge difference and Auger parameter shifts suggests that $\Delta\alpha$ may serve as an effective experimental proxy for core-hole screening strength in the material, bypassing the need for explicit charge–density difference calculations. Employing the Auger parameter analysis could thus improve the discrimination of chemically similar environments in non–metallic systems, offering a practical route to reconcile modeling and measurement.

\subsection{Auger parameter shift analysis with electrostatic models}
\label{SubSec:Full model}

In our previous work \cite{Gramatte2025}, we used a simplified electrostatic model originally developed for highly symmetric crystalline materials by Moretti to interpret aluminum Auger parameter shifts in amorphous alumina \cite{Moretti1998}. Assuming that all hydrogen atoms incorporated in the oxide are present as hydroxyl groups, the model successfully reproduced the experimental increase of \(\Delta\alpha_{\text{Al}}\) with decreasing H content (i.e., with increasing ALD growth temperature). Only for the highest ALD growth temperature of \textit{T} = 200 °C, a distinct outlier emerged, which was attributed to the gradual transformation of hydroxyl species into O ligands and interstitial (i.e. more mobile) protons and/or O$-$H$\cdots$O bridging configurations \cite{Belonoshko2004, Guo2019, Sundar2022} (which modifies the respective O ligand polarizability). Although the simplified model matched the experimental data well, it relied on the unvalidated assumption that a framework derived from crystalline environments could adequately represent the structural and chemical heterogeneity of an amorphous material. Rigorous validation against the more comprehensive model introduced in Sec. \ref{subsub:full_model_formulation} was omitted in our previous work and is addressed in this work.

\begin{figure}[H]
    \centering
    \includegraphics[width=\textwidth]{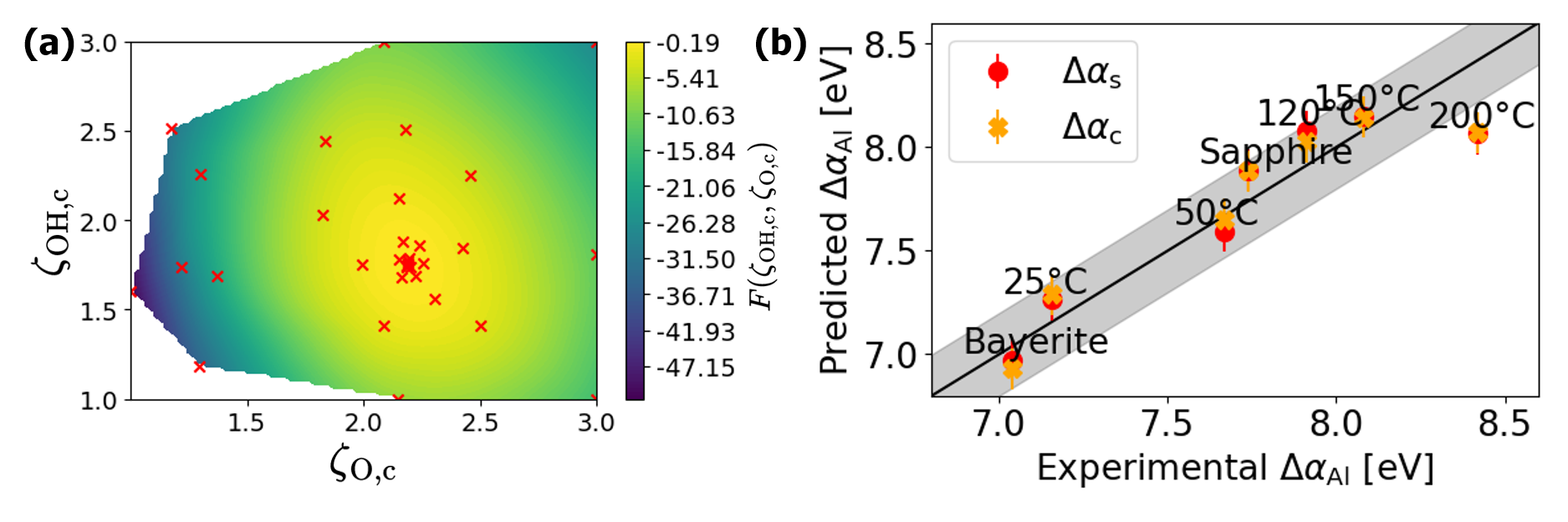}
\caption{Visualization of the Bayesian optimization process: (a) Application to the full electrostatic model, involving 34 optimization steps. (b) The shift of the Al Auger Parameter as calculated using the simple electrostatic model (\(\Delta\alpha_{\text{Al,s}}\)) and the full electrostatic model (\(\Delta\alpha_{\text{Al,c}}\)) versus the experimentally determined value, \(\Delta\alpha_{\rm Al}\). Points lying on the diagonal black line represent perfect agreement with experimental data. The surrounding gray area delineates the empirical experimental error of \(\pm 0.2\) eV.}
\label{fig:full_model_AP}
\end{figure}

The Bayesian optimization process for the complete model is considerably more computationally intensive than that for the simplified model. This is due to the necessity of calculating the distribution for \(\Delta\alpha_{\text{Al}}\) for each Al atom and frame of the MD trajectory on each occasion that a new set of ligand polarizabilities is tested. Consequently, fewer optimization steps (34) were performed, while setting the lower and upper limits in a narrower range, as the expected \(\zeta\) values lie between 1 and 3. Fig. \ref{fig:full_model_AP}(a) displays the results of Bayesian optimization (denoted by subscript letter \textit{c}), with the maximum of the objective function being achieved with \(\zeta_{\text{OH,c}} = 1.749\) \AA$^3$ and \(\zeta_{\text{O,c}} = 2.189\) \AA$^3$, which are only about 0.2\% and 1.7\% smaller than the corresponding ligand polarizability values derived using the simplified model (denoted by subscript letter \textit{s}). As discussed in Ref. \cite{Gramatte2025}, the covalent bonding of O with H replacing an ionic bond with Al in ...-Al-O-Al-... network results in the lower valence charge on O ions (as indicated by Bader charge analysis), thereby decreasing its extra-atomic screening efficiency (i.e., OH ligand polarizability with respect to the O ligand). 

The overall value of the objective function given by Eq. (\ref{f_function}) is higher than that of the simplified model, indicating that the full model provides incrementally better agreement with the experimental results. This improvement is visualized in Fig. \ref{fig:full_model_AP}(b), where \(\Delta\alpha_{\text{Al,c}}\) and \(\Delta\alpha_{\text{Al,s}}\) are compared with the experimental data, \(\Delta\alpha_{\text{Al}}\). The obtained polarizabilities fall into the middle of the accepted range for O-based ligands, from 0.9 \AA$^3$ and 3.2 \AA$^3$ \cite{Tessman1949}, and align closely with estimations of Filippone and Moretti from the analysis of Al Auger parameters for zeolite surfaces \cite{Filippone1998}.
Thus, \textit{Assumptions \#3 and \#4} in the definition of the simplified model (suggesting the use of time-and-structure-averaged descriptors of local atomic environment) are fully validated with respect to the complete model, leading to a much more straightforward interpretation of experimental Auger parameter shifts in amorphous oxides.

\begin{figure}[H]
    \centering
    \includegraphics[width=\textwidth]{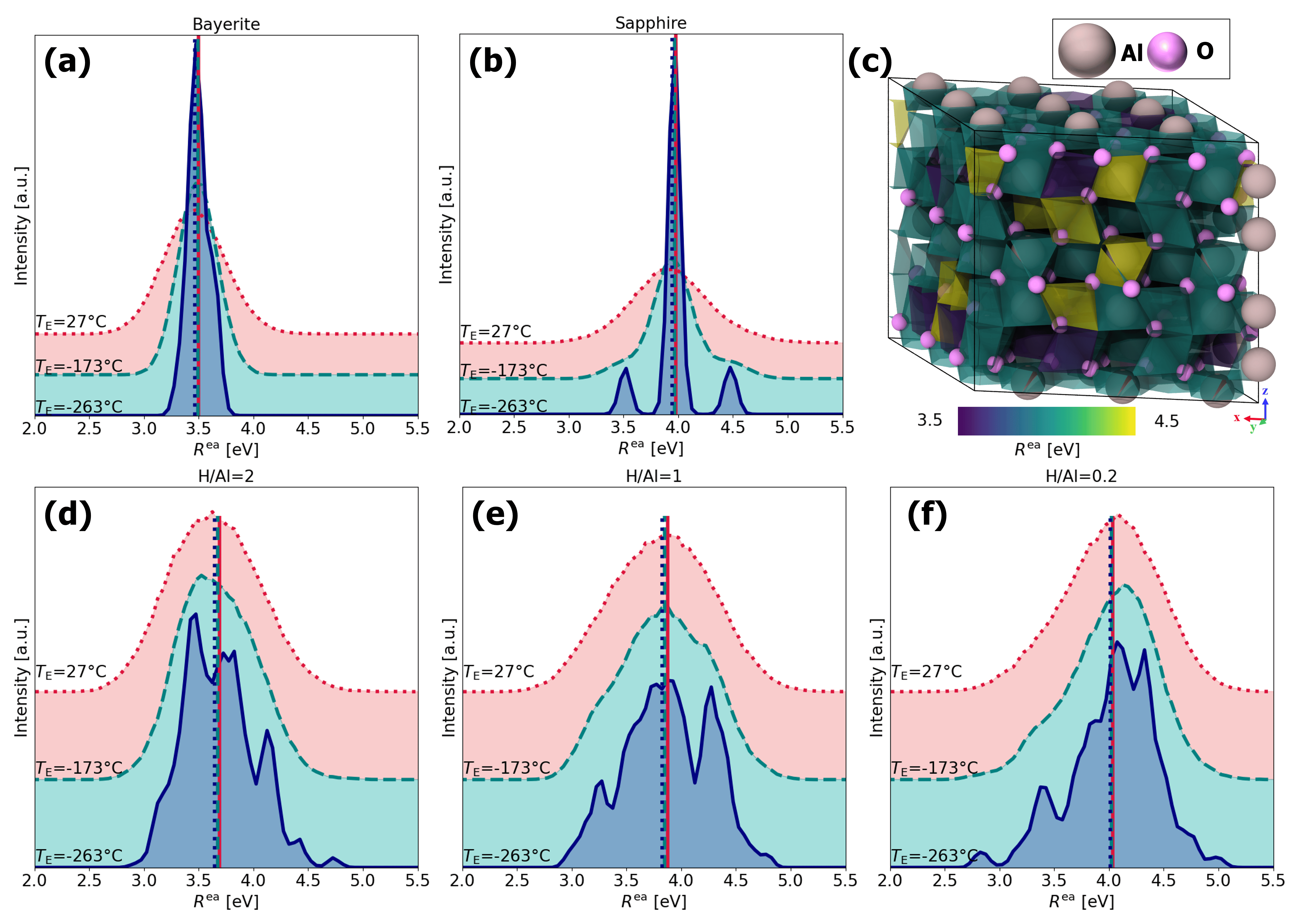}
    \caption{Distributions of the extra atomic relaxation energy, \(R^{\mathrm{ea}}\), as derived from the full model for (a) bayerite, (b) sapphire, (d) the amorphous alumina polymorph with H/Al=2 at \textit{T} = 25 °C, (e) the amorphous alumina polymorph with H/Al=1 at \textit{T} = 50 °C, and (f) the amorphous alumina polymorph with H/Al=0.2 at \textit{T} = 120 °C. The distributions are shown for three different final thermal equilibration temperatures $T_{\text{E}}$ with the solid, dashed, and dotted lines referring to \(T_{\text{E}}=27\) $^\circ$C, \(T_{\text{E}}=-173\) $^\circ$C, and \(T_{\text{E}}=-263\) $^\circ$C, respectively. The corresponding vertical lines indicate the mean value of each distribution. (c) Illustration of the building blocks corresponding to the different peaks in the \(R^{\mathrm{ea}}\) distribution plot for sapphire at \(T=-263\) $^\circ$C in (b), as visualized by colored coordination polyhedra.}
    \label{fig:Peak_splitting}
\end{figure}

To illustrate the strengths of the complete model, a detailed analysis of the atomistic simulation data is provided below. Evidently, the distribution of different types of [Al(O)$_{n-m}$(OH)$_{m}$] NNCS (with $m \leq n$ and $4\leq n \leq 6$) varies with the H content, as well as with (simulation) time. This structural diversity results in a characteristic distribution of final-state relaxation energies,  $R^{\mathrm{ea}}$, as illustrated in Fig. \ref{fig:Peak_splitting} for bayerite, sapphire and the amorphous alumina polymorphs with H/Al-ratios of 2, 1 and 0.2 after final thermal equilibration at $T_{\mathrm{E}}$ = 27$^\circ$C, -173$^\circ$C and -263$^\circ$C. 

As expected, the $ R^{\mathrm{ea}}$ distributions are symmetric and much narrower and smoother for the crystalline reference phases sapphire and bayerite (as compared to the amorphous alumina polymorphs), owing to their defined lattice periodicity and limited structural diversity (i.e., all Al cations in 6-fold coordination). The broadening of the distributions with an increasing equilibration temperature for sapphire and bayerite can be attributed to increasing thermal fluctuations. Interestingly, the $R^{\mathrm{ea}}$ distribution for sapphire reveals a symmetric fine structure at cryogenic temperatures of \(T=-263\)$^\circ$C, which originates from two specific Al-O bond lengths in sapphire \cite{Finger1978} in three distinct types of building blocks, as illustrated in Fig. \ref{fig:Peak_splitting}(c). Should the experimental resolution prove sufficient, XPS of sapphire at cryogenic temperatures may resolve such subtle differences. This suggests the broader application of cryo-XPS/HAXPES for chemical state analysis. On the contrary, the $R^{\mathrm{ea}}$ distributions for the amorphous polymorphs do not narrow with decreasing equilibration temperature, which indicates that the remaining spread in $R^{\mathrm{ea}}$ at \(T=-263\)$^\circ$C originates from structural diversity rather than from thermal fluctuations. At $T=-173$ $^\circ$C, a skewness of the $ R^{\mathrm{ea}}$ distributions becomes apparent, which are in opposite directions with respect to the mean $ R^{\mathrm{ea}}$ value for H/Al=2 and H/Al=0.2 (with H/Al=1 lying in between). As shown in Supplementary Figure S2, the resolved fine structures at $T=-173$ $^\circ$C directly relate to the distribution of (distorted) 4-fold, 5-fold, and 6-fold [Al(O)$_{n-m}$(OH)$_{m}$] NNCS.

\subsection{Evolution of short-range in amorphous alumina as a function of H content}
\label{App:chemicalEnv}

\begin{figure}[H]
    \centering
    \includegraphics[width=\textwidth]{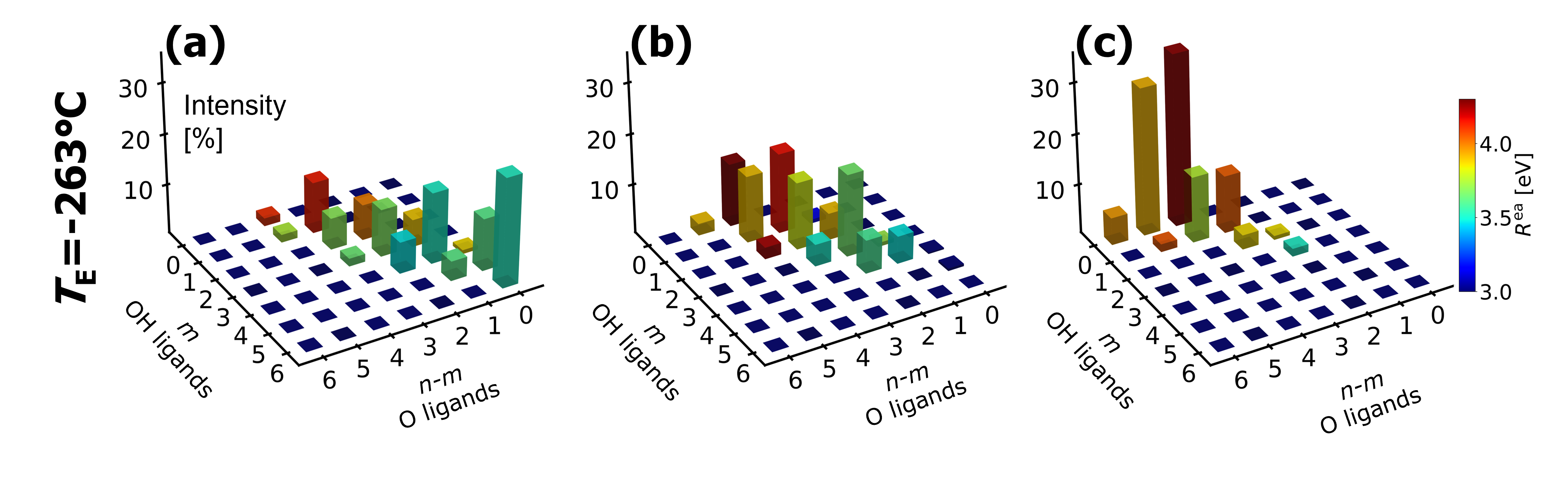}
        \caption{Top: Analysis of [Al(O)$_{n-m}$(OH)$_{m}$] NNCS in the amorphous ALD polymorphs with variable H content, as categorized based on their number of O and OH ligands for \(n \le 6\) with \(m \le n\), adopting a cutoff distance of 1.2 \AA. This study spans the entire trajectory of amorphous polymorphs maintained at \(T_{\text{E}}=-263\)$^\circ$C, as depicted in 3D histograms for three representative polymorphs with H/Al ratios of 2, 1 and 0.2 (corresponding to ALD temperatures of 25°C, 50°C, and 120°C, respectively): see panels (a), (b), and (c), respectively. The height of each histogram bin indicates the relative contribution of a given [Al(O)$_{n-m}$(OH)$_{m}$] NNCS variant to the total (integrated) \(R^{\mathrm{ea}}\) value.\\
        }
    \label{fig:Ap_contribution}
\end{figure}

In the following, specific features in the fine structure of the $R^{\mathrm{ea}}$ distribution for the amorphous polymorphs with variable H content (at $T=-263$ $^\circ$C, i.e. after lowering thermal noise) are related to the corresponding distribution of (distorted) 4-fold, 5-fold, and 6-fold [Al(O)$_{n-m}$(OH)$_{m}$] NNCS. As such, changes in short-range order of the amorphous alumina thin films as a function of their H content may be disclosed, which correlate with H-induced changes of their functional properties, such as electrical, mechanical, optical, and (H-diffusion) barrier properties \cite{Schneider2002, Nam2016, Frankberg2019}. As a first step, the short-range order around each Al atom in the computed amorphous [Al(O)$_{n-m}$(OH)$_{m}$] structures with H/Al ratios of 2, 1, and 0.2 was identified and categorized based on the coordination number, $n$, as well as on the corresponding number of O ligands, $(n-m)$, and the number of OH ligands, $m$ (following the same procedure used for determining $f_{\text{OH}}$ and $f_{\text{O}}$ in Sec. \ref{subsub:Bayesian_methods}). Considering that each Al atom can be coordinated by four up to six O(H) ligands (\textit{note}: for a cutoff distance of 1.2 \AA, only Al NNCS with $4\le n  \le 6$ are found), there are theoretically $7 \times 7 = 49$ different NNCS variants. This grouping was performed across the entire trajectory of the $T_{\mathrm{E}}=-263$ $^\circ$C equilibrated structures. The count of the NNCS variants and their relative contribution to the total (integrated) $R^{\mathrm{ea}}$ value is represented by the height of the bins in the 3D histograms: see Figs.~\ref{fig:Ap_contribution}(a-c). In these histograms, the $x$ and $y$ positions correspond to the $(n-m)$ and $m$ number of O and OH ligands for each NNCS variant, respectively, while the bin color reflects the corresponding averaged $R^{\mathrm{ea}}$ value. The \(R^{\mathrm{ea}}\) distributions originating from each of these NNCS variants, as well as the sum of all individual distributions, are displayed in Figs. \ref{fig:Ap_contribution}(a-c); SM Figs. S3 shows the same intensities as a 1D histogram. Simplified representations of Figs. \ref{fig:Ap_contribution}(a-c) are presented in SM Figs. S2 for $T_{\mathrm{E}}=-173$ $^\circ$C by classifying all 4-fold, 5-fold and 6-fold NNCs variants, independent of their O/OH ligand fraction.

Fig.~\ref{fig:Ap_contribution}(a) illustrates that for H/Al=2 (corresponding to the highest H content), octahedral [Al(OH)$_{6}$] NNCS (i.e., variant [\((n-m)=0\), \(m=6\)) are dominant. However, tetrahedrally ($n = 4$) and pentahedrally ($n = 5$) coordinated [Al(O)$_{n-m}$(OH)$_{m}$] NNCS are abundant as well and also seem to cluster hydroxyl groups, since corresponding variants with $m = 4,5$ are more frequent than those with $m < 4$. For a lower H-content of H/Al=1, as depicted in Fig. \ref{fig:Ap_contribution}(b), [Al(OH)$_{6}$] NNCS are no longer present. Most hydroxyl ligands are now allocated to 4- and 5-fold Al NNCS, which constitute the principal building blocks of amorphous alumina \cite{Meinhold1993,Kunath-Fandrei1995, MacKenzie1999, Lee2014, Shi2019, Davis2011, Xu2021}. For H/Al=1, the hydroxyl ligands spread out rather homogeneously over the 4- and 5-fold Al NNCS (i.e., variants with $m = 4, 5, 6$ are less frequent than those with $m < 4$). Finally, the H-poor amorphous polymorph with H/Al=0.2 in Fig. \ref{fig:Ap_contribution}(c) is also characterized by a dominance of 4- and 5-fold Al NNCS, which now preferentially only contain a single OH ligand.

It may thus be concluded that, for relatively low H contents H/Al $\le 1$, the H atoms are distributed roughly equally and homogeneously over the 4- and 5-fold Al NNCS, whereas for higher H/Al ratios hydrogen tends to cluster in 6-fold Al NNCS, indicative for a solid-state phase transformation towards an Al-hydroxide, presumably boehmite AlOOH \cite{guseva2025}. This evolution of short-range order around the Al cations with increasing H content is accompanied by a decrease of the mean value of $R^{\mathrm{ea}}$ (see Fig. \ref{fig:Ap_contribution}(a-c)) since $\zeta_{\text{OH}} < \zeta_{\text{O}}$  (see Sec. \ref{SubSec:Full model}). For relatively high H contents H/Al $> 1$, the measured decrease of the Al Auger parameter with increasing H content is not solely ruled by the overall increase of the OH/O ligand fraction, but also to a minor extent co-determined by the preferential formation of [Al(O)$_{n-m}$(OH)$_{m}$] building blocks, which might be regarded as a premature nucleation stage of Boehmite. This trend is also nicely reflected by a direct comparison of Figs. ~\ref{fig:Ap_contribution}(c) and (a), showing a substantial (gradual) increase of [Al(OH)$_{6}$] NNCS from H/Al = 0.2 to H/Al = 2, which shifts the mean value of $R^{\mathrm{ea}}$ to lower values: see also Figs. S3(a-c). Hence, as a general rule of thumb, [Al(O)$_{n-m}$(OH)$_{m}$] NNCS with an, on average, lower coordination number ($n$) and/or a higher number of OH ligands ($m$) govern the $ R^{\mathrm{ea}}$ contributions on the lower side of the mean $ R^{\mathrm{ea}}$ value (since the screening efficiency decreases with decreasing coordination number and $\zeta_{\text{OH} < \zeta_{\text{O}}}$; see Section \ref{subsub:simplifiedModelFormulation}): see Fig. S3 (a-c). Furthermore, Fig. \ref{fig:Ap_contribution} indicates that NNCS with the same average coordination number, $n$, tend to contribute to the same side of the $ R^{\mathrm{ea}}$ distribution, independent of the ligand type. The \( R^{\mathrm{ea}}\) distributions from \(n=5\) and \(n=6\) present similar mean and mode values (e.g. around 3.5 eV) for a high H content (H/Al=2) with the corresponding distribution for \(n=4\) being shifted by more than 0.5 eV (to 4.1 eV). This division between 5/6-fold and 4-fold NNCS contributions to $ R^{\mathrm{ea}}$ becomes less pronounced with decreasing H content due to a gradual broadening of all distributions (especially from 5- and 6-fold NNCS) towards higher \( R^{\mathrm{ea}}\) values (since $\zeta_{\text{O}} > \zeta_{\text{OH}}$). These findings suggest that the measured asymmetry (skewing) of the core-level photoelectron line of the respective cation (\textit{here}: Al 2p; as preferably measured by cryo-XPS to reduce thermal noise) might provide an alternative method for quantifying the hydrogen content in H-containing amorphous oxides. In this regard, it should be emphasized that asymmetric core-level peak shapes in XPS may also arise from satellite structures, excitation of vibrational modes, multi-electron excitations, and electron-hole pair creation in metallic valence bands \cite{Morgan2023}.

Finally, we have to make a note on the accompanied decrease in oxide density with increasing H content of roughly 20-25\% (see Fig.~\ref{fig:density_comp_andshift}b), which cannot be rationalized by the marginal (i.e., about 2\%) increase of the average Al-O bond length due to the formation of hydroxyl ligands with covalent O-H bonding characteristics (as reported in \cite{Gramatte2025}). Hence, the decreasing oxide density with increasing H content should originate from the insertion of free volume between the randomly interconnected network of corner-sharing [AlO$_{n}$]-polyhedra ~\cite{Snijders2005, Gramatte2022}, hinting at a repulsive interaction of neighbouring hydroxyl groups. Such a change in medium range order between the interconnected 4-, 5- and 6-fold [Al(O)$_{n-m}$(OH)$_{m}$] polyhedra due to the interaction of neighbouring hydroxyl ligands is not probed by the Al Auger parameter (it probes only short-range order), but is rather reflected in the measured oxide density.

\subsection{Phase decomposition characterization using Auger parameter shifts}

To evaluate the potential of Auger parameter shifts as a signature of phase decomposition in hydrogen-rich amorphous alumina, we simulated a thermal treatment of the H/Al = 2 polymorph with the highest H content. The simulation protocol involved heating the amorphous film from room temperature to $T_{\mathrm{A}} = 727$ $^\circ$C (1000 K) within 1 ns, followed by annealing for another 1 ns at $T_{\mathrm{A}}$. This is well above the crystallization temperature reported for amorphous alumina, and therefore enables phase transition to all possible phases\cite{Mavric2019a}. This annealing temperature of lies well above the crystallization window reported for amorphous Al$_2$O$_3$ (650–800 °C), thereby enabling all relevant phase transitions within the simulation time-frame \cite{Mavric2019a}. During the simulation, the simulation cell was allowed to relax in the $z$-direction, mimicking the relaxation behavior of a large-area ALD-grown film in the direction of the free surface. This annealing temperature lies well above the onset temperature for the crystallization of amorphous alumina in the range of 180 - 200 °C with $\gamma$-Al$_{2}$O$_{3}$ as the competing crystalline oxide phase in the phase transition sequence towards $\alpha$-Al$_{2}$O$_{3}$ \cite{Jeurgens2000, Snijders2002, Reichel2008a}. However, the crystallization kinetics of amorphous alumina at 200 °C are very slow (minutes to hours) with respect to the limited timescales (nano- to microseconds) of our molecular dynamics simulations, which rationalizes the much higher annealing temperature of $T_{\mathrm{A}} = 727$ $^\circ$C chosen in this study.

\begin{figure}[H]
    \centering
    \includegraphics[width=1\linewidth]{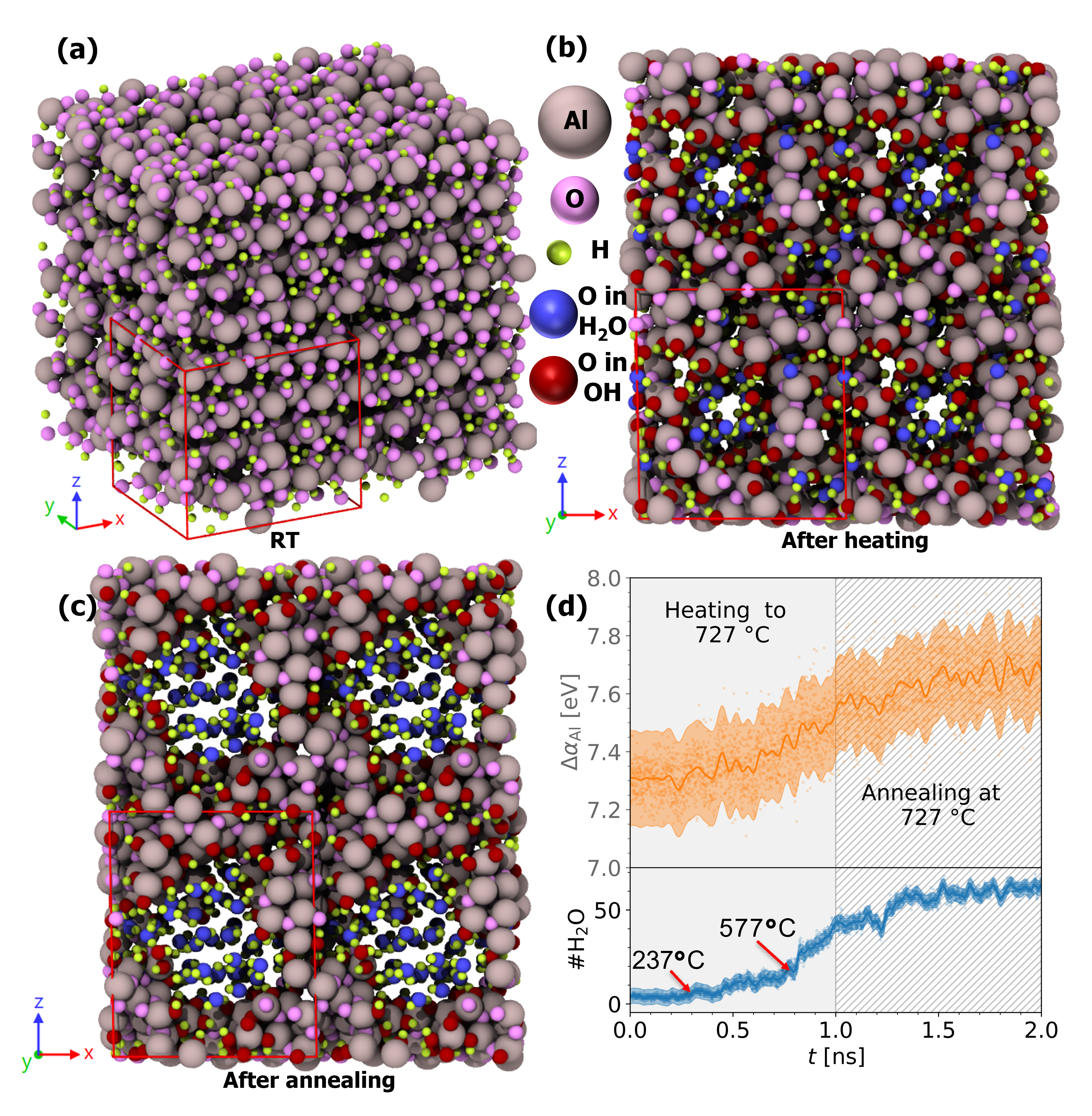}
    \caption{(a) Structure of amorphous alumina with H/Al = 2 at 25°C. The original simulation cell is shown in red, and the replicated $2x2x2$ supercell is given without edges.
(b) Structure of the same H/Al = 2 sample after heating over 1 ns to $T_{\mathrm{A}}= $727°C.
(c) Structure of the same sample after annealing for 1 ns at $T_{\mathrm{A}}$.
(d) Top: Evolution of the Al Auger parameter shift $\Delta \alpha_{\text{Al}}$ (orange) during heating (gray background) and annealing (gray hatched region) with the PFP+D3 potential. Data points from every simulation step are shown as dots. The solid line represents the mean predicted by a Gaussian process model, with the uncertainty shown as a transparent band.
Bottom: Evolution of the number of water molecules ($\mathrm{H}_2\mathrm{O}$, blue) during heating (gray background) and annealing (gray hatched region). Data points are shown as dots. The solid line represents the Gaussian process model mean, with the uncertainty visualized as a transparent region.}
    \label{fig:film_heation_gop}
\end{figure}

Figure~\ref{fig:film_heation_gop} summarizes the structural and chemical evolution during this thermal protocol. Panel (a) shows the initial structure at 25$^\circ$C, characterized by a dense hydrogenated amorphous network. After heating to 727 $^\circ$C panel (b), significant structural reorganization is visible, indicating the formation of water bubbles inside the film. Further annealing for 1 ns at 727 $^\circ$C panel (c) results in a more pronounced bubble build-up, with clear void formation and clustering of H$_2$O molecules.

Panel (d) of Figure~\ref{fig:film_heation_gop} quantifies the evolution of two key observables: the shift in the aluminum Auger parameter, $\Delta\alpha_{\text{Al}}$, as predicted by the complete electrostatic model, as well as the number of water molecules (\#$\mathrm{H}_2\mathrm{O}$) formed during heating and annealing. During the initial heating phase (gray background), $\Delta \alpha_{\text{Al}}$ exhibits a continuous increase of approximately 0.3~eV, since hydroxyl ligands are replaced by O ligands and $\zeta_{\text{O}} > \zeta_{\text{OH}}$. This shift, while subtle, lies well within the detection range of state-of-the-art XPS instruments, suggesting the feasibility of experimental in-situ monitoring of such phase transitions via Al AP tracking. The annealing stage (gray hatched region) shows an approximate stabilization of the AP signal, implying a thermodynamically metastable state in the bulk phase transition sequence towards $\gamma$-Al$_{2}$O$_{3}$ and eventually $\alpha$-Al$_{2}$O$_{3}$.

A closer look at the \#$\mathrm{H}_2\mathrm{O}$ evolution during heating reveals a first noticeable rise at around 237~$^\circ$C, marked by a red arrow in the plot. This temperature corresponds to the point at which the first water molecules can stabilize within the amorphous matrix. From that point onward, water molecules accumulate gradually. At around 577~$^\circ$C, identified as a second inflection point, the increasing number of water molecules and enhanced molecular mobility facilitate the formation of cavities and water bubbles within the film. 
This transition is visually supported by the structural evolution shown in panels (b) and (c), where water-filled voids expand noticeably from the heating to the annealing stage. After 577~$^\circ$C, the generation of water molecules accelerates, coinciding with void growth. During annealing, the number of water molecules stabilizes, possibly due to the formation of void-stabilizing OH groups and a dehydrated amorphous alumina scaffold.

George et al. \cite{George2020} observed similar nanoscale porosity development in ultrathin ALD-grown alumina films upon heating, attributing the onset of pore formation to hydroxyl group loss at comparable temperatures. Their experimental correlation between pore volume evolution and hydroxyl content provides strong supporting evidence for our atomistic findings of cavity nucleation driven by water stabilization and enhanced mobility beyond 577$^\circ$C, which would be in alignment with a polycondensation of a hydrogen-rich alumina phase to alumina and water.  

Several other studies have experimentally investigated the thermal behavior of amorphous ALD-grown alumina, particularly for films deposited under low-hydrogen conditions, such as those produced via plasma-enhanced ALD or at elevated substrate temperatures \cite{Li2013,Vermang2012,Beldarrain2012}. Upon annealing, these films consistently exhibit blistering and/or delamination. These effects are commonly attributed to hydrogen segregation—either toward the Si/Al$_2$O$_3$ interface, as a result of post-deposition thermal activation, or due to in-plane tensile stress introduced during film growth. Such segregation leads to localized gas pressure buildup beneath the dense oxide, eventually causing mechanical failure of the layer. 

This behavior contrasts with our findings on hydrogen-\textit{rich} amorphous alumina, where the porous network facilitates gradual water molecule stabilization and redistribution. In this case, the film acts not as a barrier but rather as a molecular sieve, allowing for internal reorganization without catastrophic delamination. The difference underscores the role of film morphology and hydrogen content: while dense ALD alumina layers form effective gas diffusion barriers — even for small species like H$_2$ H$_2$O, while hydrogen-rich structures can relieve internal pressure via nanoscopic void/channel formation, enabling a fundamentally different thermal decomposition pathway. 

Together, these results demonstrate that shifts in the Al Auger parameter provide a sensitive, computationally and experimentally accessible marker for detecting and characterizing phase decomposition pathways in hydrogenated amorphous alumina under thermal treatment.

\section{Conclusions}

In this work, we employed a combination of \textit{ab initio} core-level spectroscopy, electrostatic modeling, and experimental Auger parameter analysis to investigate the chemical environment and phase behavior of amorphous alumina with varying H-content, as well as crystalline reference phases. The combined theoretical and experimental approach enabled us to clarify the roles of screening, ligand coordination, and polarizability in shaping XPS results.

\begin{itemize}
    \item \textbf{Binding energy predictions with \textit{ab initio} methods}:

We demonstrated that the $\Delta$KS DFT approach, implemented via an AiiDA-XPS workflow, captures the range of core-level binding energies in aluminium-based materials. However, it does not reproduce the experimental trends across all core levels, indicating that further methodological refinement is needed to resolve closely related insulating environments with quantitative accuracy. Final-state electron density differences and radial charge profiles emphasised the importance of screening and orbital relaxation within the first neighbor shell of the core-ionized atom. Metals exhibit spherical, isotropic screening, whereas oxides and hydroxides demonstrate ligand-directed, anisotropic screening. Integrated charge displacement correlates with Auger parameter shifts and could serve as a reliable indicator of local screening strength.

\item \textbf{Auger parameter shift analysis with electrostatic models:}

Auger parameter shifts of the Al cations in Al-O-based compounds are governed by the final-state extra-atomic relaxation energy, validating the use of electrostatic models for their quantitative interpretation. Using a simplified Moretti model, we derived average ligand polarizabilities from simulated structures, confirming their applicability across varying hydrogen contents. The simplified model closely matched experimental Auger parameter shifts, and its results were consistent with those from the complete model, reinforcing the robustness and applicability of the simplified model to amorphous structures.

\item \textbf{Chemical environment of Al:}

Applying the complete electrostatic model to amorphous structures revealed thermal effects on the distribution of extra-atomic relaxation energies. Cryogenic XPS is suggested as a strategy to reduce the thermal noise and enhance understanding of local bonding.
The distributions showed a shift in mean relaxation energy and skewness with decreasing hydrogen, linked to lower Al ligand coordination and higher O ligand fractions in mixed coordination environments, [Al(O)$_{n-m}$(OH)$_{m}$] NNCS, with $n=4,5,6$.

\item \textbf{Phase decomposition characterization potential using Auger parameter shifts:}

Annealing simulations confirmed that the Al Auger parameter is a sensitive indicator of phase decomposition in H-rich amorphous alumina. A 0.3~eV rise in $\Delta\alpha_{\mathrm{Al}}$ aligned with the emergence and stabilization of H$_2$O gas bubbles.
Combining atomistic trajectories with Gaussian process regression showed that in-situ XPS monitoring could non-destructively track microstructural evolution during thermal treatment.
\end{itemize}

Overall, this work provides a detailed and predictive framework for interpreting XPS core-level shifts and Auger parameters in aluminum oxides and hydroxides. Future work should aim at validating these computational predictions through real-time experiments, improving DFT methodology for insulating environments, and extending the combined approach to other cationic systems such as Si and Mg oxides, where extra-atomic screening plays a central role.
Looking forward, experimental validation of these computational predictions will be critical: real-time XPS measurements during controlled annealing of hydrogenated alumina films should confirm the correlation between $\Delta\alpha_{\mathrm{Al}}$ and phase decomposition.

\section{Acknowledgments}
This research was supported by NCCR MARVEL, a National Centre of Competence in Research, funded by the Swiss National Science Foundation (grant number 205602).
C.C. and L.P.H.J. acknowledge financial support from the Swiss National Science Foundation (R'Equip program, Proposal No. 206021\_182987). I.T. acknowledges financial support by the Swiss National Science Foundation (SNSF) Project Funding (grant numbers 200021-227641 and 200021-236507). We acknowledge access to Alps at the Swiss National Supercomputing Centre, Switzerland, under the MARVEL's share with the project ID mr30. All authors thank Prof. Giuliano Moretti and Prof. Patrik Hoffmann for the fruitful exchange and helpful suggestions during manuscript preparation.

\section{Data availability statement}
Data supporting the findings of this study are openly available at the following URL/DOI: \url{https://doi.org/10.24435/materialscloud:9v-61} \cite{Gramatte_Auger_data}.

\bibliographystyle{ieeetr} 
\bibliography{PhD}

\end{document}